\documentclass[aps,  twocolumn,superscriptaddress, longbibliography]{revtex4-2}

\usepackage{setspace}
\usepackage{lipsum,adjustbox}
\usepackage[hidelinks,colorlinks=true,linkcolor=blue,citecolor=blue]{hyperref}
\usepackage{amsmath}
\usepackage{amssymb}
\usepackage{mathrsfs}
\usepackage{verbatim}
\usepackage{epsfig}
\usepackage{color}
\usepackage{soul}
\usepackage{setspace}
\usepackage{graphicx}
\usepackage{physics}
\usepackage{graphicx}
\usepackage{mathtools}
\usepackage{algorithm}
\usepackage{algpseudocode}
\usepackage{setspace}
\usepackage{enumerate}

\makeatletter
\renewcommand{\fnum@algorithm}{\fname@algorithm}
\makeatother
\algnewcommand{\algorithmicgoto}{\textbf{go to}}%
\algnewcommand{\Goto}[1]{\algorithmicgoto~\ref{#1}}%
\graphicspath{{.}{./EPS/}}

\newcommand{\hbrho}{\boldsymbol{\rho}}
\newcommand{\hbsigma}{\boldsymbol{\sigma}}

\usepackage[margin=1.5cm]{geometry}

\usepackage{graphicx}

\begin{document}
\title{Statistical evaluation and optimization of entanglement purification protocols}
\author{Francesco Preti}
\affiliation{Forschungszentrum J\"ulich, Institute of Quantum Control (PGI-8), D-52425 J\"ulich, Germany}
\affiliation{Institute for Theoretical Physics, University of Cologne, D-50937 K\"oln, Germany}
\author{J\'ozsef Zsolt Bern\'ad}
\affiliation{Forschungszentrum J\"ulich, Institute of Quantum Control (PGI-8), D-52425 J\"ulich, Germany}

\date{\today}

\begin{abstract}
Quantitative characterization of two-qubit entanglement purification protocols is introduced. Our approach is based on the concurrence and the hit-and-run algorithm applied to the convex set of all two-qubit states. We demonstrate that pioneering protocols are unable to improve the estimated initial average concurrence of almost uniformly sampled density matrices, however, as it is known, they still generate pairs of qubits in a state that is close to a Bell state. We also develop a more efficient protocol and investigate it numerically together with a recent proposal based on an entangling rank-$2$ projector. Furthermore, we present a class of variational purification protocols with continuous parameters and optimize their output concurrence. These optimized algorithms turn out to surpass former proposals and our protocol by means of not wasting too many entangled states.
\end{abstract}

\maketitle

\section{Introduction}
\label{sec:I}

Entanglement purification protocols aim to overcome the destructive effects of non-ideal channels by generating highly entangled states from a large number of noisy entangled states \cite{Duer2007}. In this approach, the almost perfectly entangled pairs obtained are used in quantum teleportation \cite{Bennett1993}, and thus quantum data can be transmitted across the channels. The other possible solution to this problem is to use quantum error correction \cite{Devitt}, when quantum data is sent through the channel by adding enough redundancy, e.g., increasing the number of qubits, such that the original information is recoverable even in the presence of noise. While in quantum error correction the sources of errors and their models have to be identified, the first entanglement purification protocols \cite{Bennett1996, Deutsch} offer solutions for general noisy two-qubit states. However, these protocols convert a general two-qubit state to either Werner or Bell diagonal states by using local random transformations, which waste useful entanglement \cite{Bennett2, Horodecki}. This was found by investigating particular entangled states, whose entanglement is destroyed in the first step of the protocol, and this issue was also confirmed in a different entanglement purification approach \cite{Torres}. A quantitative characterization of the ratio between wasted and improved entangled states is missing and this paper is devoted to investigating this question. 

The quest for optimality is ongoing and important research in quantum physics ranging from experimental protocols to algorithms. In most cases, numerical optimization based on continuous and discrete parameters
is carried out to enhance performances with respect to different figures of merit. This has recently led to the improvement of entanglement generation \cite{Melnikov1221}, unitary compilation and state preparation \cite{preti2024hybrid}, quantum error correction \cite{Martinez}, communication \cite{Nautrup2019optimizingquantum, Wallnfer2019MachineLF}, and algorithms \cite{Wecker}, which are only a few examples of the vast literature. Concerning entanglement purification, discrete optimizations have been considered  for Werner states \cite{Krastanov2019optimized} and we have started to investigate continuous optimization for certain families of states \cite{Preti}. Therefore, in this paper we not only evaluate some existing proposals but also search for more optimal protocols with a lower amount of wasted entangled states.

In this study, we avail ourselves of the hit-and-run algorithm to generate asymptotically and effectively uniformly distributed two-qubit density matrices \cite{Sauer}. It is known that approximately $24.24$ \% of the generated two-qubit states are separable, a numerical result that has also been confirmed by other methods \cite{Dunkl,Shang,Strunz,Fei}. Therefore, almost one-fourth of the states are immediately useless in an entanglement purification protocol. In the case of entangled states, we use the concurrence \cite{Wootters} to measure the improvement or deterioration induced by a protocol. The choice of the concurrence and its advantage over the usually employed fidelity will be explained in the context of those protocols, which have two Bell states as stable fixed points. We calculate the average concurrence over the whole sample, which will be our cost function. Any change of this cost function characterizes only the whole sample of two-qubit states, whereas the concurrence of individual states may show different behaviors. The obtained estimates allow us to compare quantitatively some existing entanglement purification protocols and to search numerically for more optimal scenarios. These scenarios consist of the use of the $SU(4)\times SU(4)$ group to find optimal locally entangling gates. We assume throughout the whole paper that these operations can be performed without errors on both sides of the noisy channel.

Our search for optimal protocols is based on the limited memory Broyden-Fletcher-Goldfarb-Shanno (LBFGSB) optimization algorithm with bounds \cite{Nocedal, Byrd}. To use this method we employ the Euler angle parametrization of the $SU(4)\times SU(4)$ group \cite{Tilma1}, where the group is constructed from angles, which form a hyperrectangle in a $30$-dimensional Euclidean vector spaces. Thus, every vector represents an element of the group and its neighborhood is defined by the Euclidean norm. The resulting parametrization allows us to construct a variational cost function that depends on a unitary in $SU(4) \times SU(4)$ and that can be minimized with the LBFGSB method due to automatic differentiation \cite{jax2018github}. With this strategy, we demonstrate that it is possible to increase the percentage of entangled states purified by computer-designed protocol and achieve better performances. Furthermore, our approach introduces a systematic and general way to compare different proposals for entanglement purification protocols, where more realistic experimental considerations can be incorporated upon request.

The paper is organized as follows. In Sec. \ref{sec:II}, we recall the mathematical description of some known entanglement purification protocols. Furthermore, we present a controlled-NOT(CNOT) gate-based protocol and also our variational approach. A brief description of the numerical approach is given in Sec. \ref{sec:III}. Numerical results are presented and discussed in Sec. \ref{sec:IV}. Section \ref{sec:V} contains a summary and our conclusions. Some technical details are provided in the Appendices.

\section{Entanglement purification}
\label{sec:II}

Entanglement purification describes a protocol between two nodes of a quantum network with the task of extracting highly entangled states, e.g., Bell states, from arbitrarily entangled states. There are several categories of purification protocols, based on the way the entangled states are distilled: filtering, recurrence, hashing, and breeding protocols \cite{Duer2007}. Here, our focus lies on bipartite recurrence protocols, where the two nodes $A$ and $B$ initially share a pair of identical two-qubit states: 
\begin{align}
\label{eq:identicalshare}
    \varrho = \rho^{A_1, B_1} \otimes \rho^{A_2,B_2}.
\end{align}
The goal is to increase the entanglement of one of the pairs by performing local entangling operations and measurements in the nodes. Finally, classical communication between $A$ and $B$ is used. In this paper, we consider two-qubit states in the representation:
\begin{equation}
\rho=\sum^4_{i,j=1} r_{ij} \ket{i} \bra{j}, 
\label{eq:representation}
\end{equation}
with the Bell states
\begin{eqnarray}\label{eq:bellstates}
  \ket{1}&=&\tfrac{1}{\sqrt2}\left(\ket{01}-\ket{10}\right),
  \quad 
  \ket{2}=\tfrac{1}{\sqrt2}\left(\ket{01}+\ket{10}\right), \nonumber \\
  \ket{3}&=&\tfrac{1}{\sqrt2}\left(\ket{00}- \ket{11}\right), \quad
  \ket{4}=\tfrac{1}{\sqrt2}\left(\ket{00}+\ket{11}\right). \nonumber
\end{eqnarray}
Furthermore, we use the notation $ r_{j}= r_{jj}$. The properties $\mathrm{Tr}\{\rho\}=1$ and $\rho^\dagger=\rho$ yield the following relations:
\begin{eqnarray}
 r_1+r_2+r_3+r_4=1,\quad r_{ij}=\left(r_{ji}\right)^*. \label{eq:dprop}
\end{eqnarray}

In the following, we give a short overview of three purification protocols and introduce a CNOT-based approach and our variational method for the search of more optimal strategies.

\subsection{Bennett protocol}
\label{sec:IIA}

The seminal protocol introduced in Ref.~\cite{Bennett1996} is based on the CNOT gate and allows one to distill the Bell state $\ket{1}$ from a large ensemble of initial two-qubit states $\rho$. In fact, the state $\ket{1}$ is only reached in the asymptotic limit of the purification rounds. The protocol operates only on Werner states \cite{Werner}
\begin{equation}
    \rho_W = r_1 \ket{1}\bra{1} + \frac{1-r_1}{3} (I_4 - \ket{1}\bra{1})
\end{equation}
where $I_4$ is the $4 \times 4$ identity matrix. The transformation of a general state $\rho$ of Eq.~\eqref{eq:representation} into a Werner state $\rho_W$ can be achieved by local random unitary rotations \cite{Bennett2}, i.e., the so-called twirling operation, which is given by
\begin{equation}\label{eq:rhotowerner}
    \rho_W = \frac{1}{12}  \sum_{j=1}^3 K^\dagger_j \left(\sum_{i=1}^4  K_{i}^{\dagger}K_{i}^{\dagger} \rho K_{i} K_{i} \right) K_j
\end{equation}
with the transformations
\begin{eqnarray}
  K_j&=&u_j^A\otimes u_j^B,\quad
  u_1=\frac{I_2+i\sigma_x}{\sqrt2},\quad
  u_2=\frac{I_2-i\sigma_y}{\sqrt2},\quad
  \nonumber\\
   u_3&=&i\ket{0}\bra{0}+\ket{1}\bra{1},\quad  u_4=I_2,
  \label{Ks}
\end{eqnarray}
the Pauli matrices $\sigma_x$ and $\sigma_y$,
and the $2 \times 2$ identity matrix $I_2$. The protocol consists of the following steps:
\begin{enumerate}[(i)]
    \item Bring the state $\rho$ into Werner form by using Eq.~\eqref{eq:rhotowerner}.
    \item Apply local $\sigma_y$ rotations on qubits $A_1$ and $A_2$.
    \item Perform the bilateral operation  $ U_{\rm CNOT}^{A_1\rightarrow A_2}\otimes U_{\rm CNOT}^{B_1\rightarrow B_2}$.
    \item Measure the target pair ($A_2,B_2$) in the eigenbasis of the Pauli matrix $\sigma_z$ with corresponding results $(m,n)$,  where $m,n \in \{0,1\}$. Keep the pair ($A_1,B_1$) if the measurement result is either $m=n=0$ or $m=n=1$ and finally perform a $\sigma_y$ rotation on $A_1$.
\end{enumerate}
An elementary step of the protocol yields $\rho'_W$ with
\begin{equation}
r'_1=\frac{1 - 2 r_1 + 10 r^2_1}{5 - 4 r_1 + 8 r^2_1}
\label{eq:Bennettit}
\end{equation}
and success probability $P_s=(5 - 4 r_1 + 8 r^2_1)/9$. 
The state $\rho'_W$ becomes more entangled than $\rho_W$ if
\begin{equation}
    2 r_1-1>0. \label{eq:Bennettcond}
\end{equation}

\subsection{Deutsch protocol}
\label{sec:IIB}

This protocol is conceptually similar to the previous one and operates on Bell diagonal states \cite{Deutsch}
\begin{equation}
\rho_B=\sum^4_{i=1} r_i \ket{i} \bra{i}.
\end{equation}
The transformation of a general state $\rho$ of Eq.~\eqref{eq:representation} into a Bell diagonal state reads
\begin{equation}
 \rho_B = \frac{1}{4} \sum_{i=1}^4  K_{i}^{\dagger}K_{i}^{\dagger} \rho K_{i} K_{i},
\end{equation}
where the $K_i$ are given in Eq. \eqref{Ks}. The Deutsch protocol can be summarized in three steps:
\begin{enumerate}[(i)]
    \item Apply the unitary operation 
$ u_1^{\dagger A_1}\otimes u_1^{\dagger A_2}\otimes u_1^{B_1}\otimes u_1^{B_2}$, see Eq. \eqref{Ks}.
\item Perform the bilateral operation  $\hat U_{\rm CNOT}^{A_1\rightarrow A_2}\otimes\hat U_{\rm CNOT}^{B_1\rightarrow B_2}$.
\item Measure the target pair ($A_2,B_2$) in eigenbasis of $\sigma_z$ with corresponding results $(m,n)$,  where $m,n \in \{0,1\}$. Keep the pair ($A_1,B_1$) if the measurement result is either $m=n=0$ or $m=n=1$.

\end{enumerate}
After applying the purification protocol we obtain a new Bell diagonal state $\rho'_B$, which is described
by the map
\begin{eqnarray}
r'_1&=&\frac{2 r_2 r_3}{C}, \quad 
r'_2=\frac{r^2_2 + r^2_3}{C}, \nonumber \\
r'_3&=&\frac{2 r_1 r_4}{C}, \quad 
r'_4=\frac{r^2_1 + r^2_4}{C}, \label{eq:Deutschit}
\end{eqnarray}
where $C=(r_1+r_4)^2+(r_2+r_3)^2=P_s$ is the success probability. Entanglement of the state $\rho'_B$ compared to $\rho_B$ is enhanced if
\begin{eqnarray}
 (2r_1-1)(1-2r_4)>0 \quad \text{or} \quad (2r_2-1)(1-2r_3)>0.
 \nonumber \\
 \label{eq:Deutschcond}
\end{eqnarray}
Depending on which one of the above conditions is fulfilled, the protocol distills asymptotically either $\ket{4}\bra{4}$ or $\ket{2}\bra{2}$ \cite{Macchiavello}, i.e, the Bell states $\ket{4}$ and $\ket{2}$.

\subsection{Matter-field interaction-based protocol}
\label{sec:IIC}

A key step in the previous protocols is the application of an entangling unitary transformation in the nodes $A$ and $B$. The motivation to choose the abstract CNOT gate comes mainly from classical computer science. However, any entangling quantum operation can serve the same purpose as shown in Ref. \cite{Bernad} for a cavity QED  setup, where an entangling transformation emerges from matter-field interactions and is modeled by a rank-$2$ projector \cite{Bernad}
\begin{equation}
M=\ket{1}\bra{1}+\ket{3}\bra{3}. \label{eq:M}
\end{equation}
The purification protocol based on this projector consists of the following steps:
\begin{enumerate}[(i)]
    \item Apply $M$ in both nodes which results in the state
    \begin{equation}
    \varrho'=
\frac{ \Pi \varrho \Pi^\dagger }{\mathrm{Tr} 
\left\{\Pi^\dagger \Pi \varrho\right\}}
,\quad  \Pi=M^{A_1,A_2} \otimes  M^{B_1,B_2}. \nonumber   
    \end{equation}
 \item Measure one of the pairs, say pair $(A_2,B_2)$, in eigenbasis of $\sigma_z$ with corresponding results $(m,n)$,  where $m,n \in \{0,1\}$. This results in the state $\rho^{A_1,B_1}_{m,n}$. 
 \item The final two-qubit state is then given by
\begin{equation}
  \rho'=
  \left(v_m^{A_1} \otimes v_{n+1}^{B_1}\right)
  \rho^{A_1,B_1}_{m,n}
  \left(v_m^{A_1} \otimes v_{n+1}^{B_1}\right)^\dagger,
  \nonumber
\end{equation}
where $v_n=\left(i\ket{0}\bra{0}+\ket{1}\bra{1}\right)\sigma^n_x$.
\end{enumerate}
The resulting state $\rho'$ depends only on seven real parameters instead of fifteen and the map of the protocol reads \begin{eqnarray}
 r'_1&=&\frac{r^2_{1}+r^2_{3}-r^2_{13}-r^2_{31}}{D}, \quad 
 r'_3=2\frac{r_{2}r_{4}-|r_{24}|^2}{D} \nonumber \\
 r'_2&=&\frac{r^2_{2}+r^2_{4}-r^2_{24}-r^2_{42}}{D}, \quad
 r'_4=2\frac{r_{1}r_{3}-|r_{13}|^2}{D}, \label{eq:MFIit} \\
 r'_{12}&=&\frac{r^2_{12}+r^2_{34}-r^2_{14}-r^2_{32}}{D},  
  \quad r'_{34}=2\frac{r_{21}r_{43}-r_{23}r_{41}}{D},
 \nonumber \\
r'_{13}&=&r'_{14}=r'_{23}=r'_{24}=0, \nonumber
\end{eqnarray}
where the success probability $P_s=D/2$ and 
\begin{equation}
D=(r_{1}+r_{3})^2+(r_{2}+r_{4})^2-(r_{13}+r_{31})^2-(r_{24}+r_{42})^2. \nonumber
\end{equation}
Due to the relations in Eq.~\eqref{eq:dprop}, we also have
$r'_{21}=\left(r'_{12}\right)^*$ and $r'_{43}=\left(r'_{34}\right)^*$. Therefore, $r'_1$, $r'_2$, $r'_3$, $r'_4$, $r'_{12}$, $r'_{21}$, $r'_{34}$, and $r'_{43}$ are the nonvanishing elements of $\rho'$. This protocol was analysed in Ref. \cite{Torres} and it was found that either 
\begin{equation}
 (2r_1-1)(1-2r_3)>-
  (2{\rm Im [r_{13}]})^2-(2{\rm Re}[r_{24}])^2 
  \label{eq:MFIcond1}
\end{equation}
or
\begin{equation}
 (2r_2-1)
  (1-2r_4)>-(2{\rm Im [r_{24}]})^2-(2{\rm Re}[r_{13}])^2 
  \label{eq:MFIcond2}
\end{equation}
is fulfilled, then $\rho'$ becomes more entangled than $\rho$. In a further iteration, the output state $\rho''$ remains in the same form as the input state $\rho'$ and according to Eq.~\eqref{eq:MFIit} its elements read
\begin{eqnarray}
 r''_1&=&\frac{r'^2_{1}+r'^2_{3}}{D'}, \quad r''_2=\frac{r'^2_{2}+r'^2_{4}}{D'}, \quad
 r''_3=2\frac{r'_{2}r'_{4}}{D'} \nonumber \\
 r''_4&=&2\frac{r'_{1}r'_{3}}{D'}, \quad  r''_{12}=\frac{r'^2_{12}+r'^2_{34}}{D'}=\left(r''_{21}\right)^*, \nonumber \\
r''_{34}&=&2\frac{r'_{21}r'_{43}}{D'}=\left(r''_{43}\right)^*,
 \nonumber
\end{eqnarray}
where
\begin{equation}
D'=(r'_{1}+r'_{3})^2+(r'_{2}+r'_{4})^2. \nonumber    
\end{equation}
In the asymptotic limit, the protocol converts all states fulfilling Eq.~\eqref{eq:MFIcond1} into $\ket{1}\bra{1}$ and Eq.~\eqref{eq:MFIcond2} into $\ket{2}\bra{2}$.

\subsection{A CNOT-based protocol}
\label{sec:IID}

The proof presented in Ref. \cite{Torres} is very general and one can apply it to obtain a better CNOT-based protocol, where the transformations into either Werner or Bell diagonal state are omitted. Our proposed protocol reads
\begin{enumerate}[(i)]
    \item Apply the unitary operation 
$ u_1^{\dagger A_1}\otimes u_1^{\dagger A_2}\otimes u_1^{B_1}\otimes u_1^{B_2}$, see Eq. \eqref{Ks}.
\item Perform the bilateral operation  $\hat U_{\rm CNOT}^{A_1\rightarrow A_2}\otimes\hat U_{\rm CNOT}^{B_1\rightarrow B_2}$.
\item Measure the target pair ($A_2,B_2$) in eigenbasis of $\sigma_z$ with corresponding results $(m,n)$,  where $m,n \in \{0,1\}$. Keep the pair ($A_1,B_1$) if the measurement result is $m=n=1$.
\end{enumerate}
The protocol is described by the map
\begin{eqnarray}
 r'_1&=&2\frac{r_{2}r_{3}-|r_{23}|^2}{E}, \quad 
 r'_2=\frac{r^2_{2}+r^2_{3}+r^2_{23}+r^2_{32}}{E} \nonumber \\
 r'_3&=&2\frac{r_{1}r_{4}-|r_{14}|^2}{E}, \quad
 r'_4=\frac{r^2_{1}+r^2_{4}+r^2_{14}+r^2_{41}}{E} , \label{eq:CNOTit}  \\
 r'_{13}&=&2\frac{r_{24}r_{31}-r_{21}r_{34}}{E},  
  \quad r'_{24}=\frac{r^2_{21}+r^2_{31}+r^2_{24}+r^2_{34}}{E},
 \nonumber \\
 r'_{12}&=&r'_{14}=r'_{23}=r'_{34}=0, \nonumber
\end{eqnarray}
where the success probability $P_s=E/2$ and 
\begin{equation}
E=(r_{1}+r_{4})^2+(r_{2}+r_{3})^2+(r_{14}-r_{41})^2+(r_{23}-r_{32})^2. \nonumber
\end{equation}
According to Eq. \eqref{eq:dprop}, we also have
$r'_{31}=\left(r'_{13}\right)^*$ and $r'_{42}=\left(r'_{24}\right)^*$. Therefore, $r'_1$, $r'_2$, $r'_3$, $r'_4$, $r'_{13}$, $r'_{31}$, $r'_{24}$, and $r'_{42}$ are the nonvanishing elements of $\rho'$.
To increase the degree of entanglement of $\rho'$, the state 
$\rho$ has to fulfill either
\begin{equation}
 (2r_1-1)(1-2r_4)>-
  (2{\rm Im [r_{23}]})^2-(2{\rm Re}[r_{14}])^2 
  \label{eq:CNOTcond1}
\end{equation}
or
\begin{equation}
 (2r_2-1)
  (1-2r_3)>-(2{\rm Im [r_{14}]})^2-(2{\rm Re}[r_{23}])^2. 
  \label{eq:CNOTcond2}
\end{equation}
It is worth noting that one can keep the pair ($A_1,B_1$) if the measurement result on ($A_2,B_2$) is $m=n=0$. However, in this case, the entanglement purification works only if either 
\begin{equation}
 (2r_1-1)(1-2r_4)>
  (2{\rm Im [r_{23}]})^2+(2{\rm Re}[r_{14}])^2 
  \label{eq:CNOTcond1extra}
\end{equation}
or
\begin{equation}
 (2r_2-1)
  (1-2r_3)>(2{\rm Im [r_{14}]})^2+(2{\rm Re}[r_{23}])^2. 
  \label{eq:CNOTcond2extra}
\end{equation}
is fulfilled. These conditions are more restrictive than their counterparts in Eqs.~\eqref{eq:CNOTcond1} and \eqref{eq:CNOTcond2}. They work for Bell diagonal states, but many states do not obey them, and therefore for a general purification strategy the pair ($A_1,B_1$) has to be discarded,
whenever the pair ($A_2,B_2$) is measured in the state $\ket{00}$. In a further iteration, the output state $\rho''$ remains in the same form as the input state $\rho'$ and according to Eq.~\eqref{eq:CNOTit} its elements read
\begin{eqnarray}
 r''_1&=&2\frac{r'_{2}r'_{3}}{E'}, \quad r''_2=\frac{r'^2_{2}+r'^2_{3}}{E'}, \quad
 r''_3=2\frac{r'_{1}r'_{4}}{E'} \nonumber \\
 r''_4&=&\frac{r'^2_{1}+r'^2_{4}}{E'}, \quad  r''_{13}=2\frac{r'_{24}r'_{31}}{E'}=\left(r''_{31}\right)^*, \nonumber \\
r''_{24}&=&\frac{r'^2_{31}+r'^2_{24}}{E'}=\left(r''_{42}\right)^*,
 \nonumber
\end{eqnarray}
where
\begin{equation}
E'=(r'_{1}+r'_{4})^2+(r'_{2}+r'_{3})^2. \nonumber    
\end{equation}
If the states fulfill the condition given in Eq.~\eqref{eq:CNOTcond1} then the protocol can distill $\ket{4}\bra{4}$, otherwise in the case of Eq.~\eqref{eq:CNOTcond2}, the state
$\ket{2}\bra{2}$ is obtained. It is worth noting that these Bell states can also be reached not only in the asymptotic limit. Based on the second example in Ref. \cite{Torres}, one can pick $r_2=c$, $r_1=r_4=(1-c)/2$, and $r_{14}=\left(r_{41}\right)^*=i(1-c)/2$ with $c\in (0,1]$ to observe that one iteration of the protocol yields $\ket{2}\bra{2}$ with success probability $c^2/2$. This state cannot be purified by the Bennett protocol. If $c\in (0.5,1]$, then the Deutsch protocol approaches the Bell state $\ket{2}$ only asymptotically. 

\subsection{Variational purification protocols}
\label{sec:IIE}

In order to implement a numerical search for optimal protocols, a cost function has to be defined. A convenient choice for the measure of the performance of a two-qubit-based entanglement purification protocol is the concurrence \cite{Preti}, for which we use the definition introduced in Ref.~\cite{Wootters}:
\begin{equation}\label{eq:concurrence}
 \mathcal{C}(\rho)=\max\{0,\lambda_1-\lambda_2-\lambda_3-\lambda_4\}.
\end{equation}
Here $\lambda_1, \lambda_2, \lambda_3, \lambda_4$ are the square roots, listed in decreasing order, of the non-negative eigenvalues of the non-Hermitian matrix
\begin{equation}
 \tilde \rho = \rho (\sigma_{y}\otimes\sigma_{y})\rho^{*}(\sigma_{y}\otimes\sigma_{y}), \nonumber
\end{equation}
where the asterisk denotes the complex conjugation in the standard basis. The advantage of the concurrence lies in the fact that it treats all maximally entangled states equivalently. As we have discussed in the previous sections, the maps there can have two stable fixed points corresponding to two Bell states and they can approach both of them depending on the properties of the input density matrix. Therefore, picking a fidelity with respect to a Bell state would not work, because the optimization would suppress the convergence towards another Bell state or any maximally entangled state, which can be reached otherwise. A different argument against the use of fidelity with an example is given in Ref. \cite{Preti}.

Let us consider a general unitary matrix $V \in SU(4) \otimes SU(4)$ acting on $A_1, A_2$ and $B_1, B_2$.
We employ the Euler-angle parametrization of $SU(4)$ \cite{Tilma1},
\begin{eqnarray} \label{eq: eulerU}
 U(\boldsymbol{\alpha})=e^{i \sigma_3 \alpha_1} e^{i \sigma_2 \alpha_2} e^{i \sigma_3 \alpha_3} e^{i \sigma_5 \alpha_4} e^{i \sigma_3 \alpha_5} e^{i \sigma_{10} \alpha_6} 
 e^{i \sigma_3 \alpha_7} e^{i \sigma_2 \alpha_8} \nonumber \\ 
 e^{i \sigma_{3} \alpha_9} e^{i \sigma_5 \alpha_{10}} e^{i \sigma_3 \alpha_{11}} e^{i \sigma_{2} \alpha_{12}} 
 e^{i \sigma_3 \alpha_{13}} e^{i \sigma_8 \alpha_{14}} e^{i \sigma_{15} \alpha_{15}}, \nonumber 
\end{eqnarray}
with $\boldsymbol{\alpha}=(\alpha_1, \alpha_2, \dots, \alpha_{15})^T \in \mathbb{R}^{15}$ ($T$ denotes the transposition), and
\begin{eqnarray}
 && 0 \leqslant \alpha_1, \alpha_3, \alpha_5, \alpha_7, \alpha_9, \alpha_{11}, \alpha_{13} \leqslant \pi, \nonumber \\
 &&  0 \leqslant \alpha_2, \alpha_4, \alpha_6, \alpha_8, \alpha_{10}, \alpha_{12} \leqslant \frac{\pi}{2} \nonumber \\
 && 0 \leqslant \alpha_{14} \leqslant \frac{\pi}{\sqrt{3}}, \quad 0 \leqslant \alpha_{15} \leqslant \frac{\pi}{\sqrt{6}}, \label{eq:anglecond}
\end{eqnarray}
where $\sigma_i,\ i=1, ..., 15$ form a Gell-Mann type basis of the Lie group $SU(4)$ (see the Appendix in Ref.~\cite{Preti}, where the same approach to angle parametrization was used). Then we have
\begin{align}
    V(\boldsymbol \alpha_{AB}) = U^{A_1, A_2}(\boldsymbol \alpha_A) \otimes U^{B_1, B_2}(\boldsymbol \alpha_B)
\end{align}
where $\boldsymbol \alpha_{AB} = (\boldsymbol \alpha_A, \boldsymbol \alpha_B)^T$ and thus $\boldsymbol \alpha_{AB} \in \mathbb{R}^{30}$. The Euler-angle parametrization guarantees that the group is not naively overcounted \cite{Tilma1}.  It is worth noting that due to the nature of the problem, one cannot consider operations that improve entanglement across the channel, i.e., general $SU(16)$ operations on $A$ and $B$, because badly entangled states are the result of the noisy channel.
 
The purification protocol for our numerical investigations reads as follows.
\begin{enumerate}[(i)]
    \item Apply entangling operations in both nodes which results in the state
    \begin{equation}
    \varrho'=
\frac{ \Pi \varrho \Pi^\dagger }{\mathrm{Tr} 
\left\{\Pi^\dagger \Pi \varrho\right\}},
 \label{eq:varprot}   
    \end{equation}
 where $\Pi$ is a quantum operation. In the numerical simulations, apart from one particular case, $\Pi$ equals $V(\boldsymbol \alpha_{AB})$.   
 \item Measure the pair ($A_2,B_2$) in eigenbasis of $\sigma_z$ with corresponding results $(m,n)$,  where $m,n \in \{0,1\}$. The four projectors of the $(m,n)$ measurement results are:
 \begin{align}
    &(0,0) \rightarrow P_{1} = I_{A_1, B_1} \otimes \ket{00}\bra{00}_{A_2, B_2},  \label{eq:P00}\\
    &(0,1) \rightarrow P_{2} = I_{A_1, B_1} \otimes \ket{01}\bra{01}_{A_2, B_2}, \label{eq:P01}\\
    &(1,0) \rightarrow P_{3} = I_{A_1, B_1} \otimes \ket{10}\bra{10}_{A_2, B_2}, \label{eq:P10}\\
    &(1,1) \rightarrow P_{4} = I_{A_1, B_1} \otimes \ket{11}\bra{11}_{A_2, B_2}. \label{eq:P11}
\end{align}
We define a selection function $\pi$ (measurement policy) that chooses which measurement result is kept:
\begin{align}
   (\rho^{A_1,B_1}_1, \rho^{A_1,B_1}_2, \rho^{A_1,B_1}_3, \rho^{A_1,B_1}_4) \mapsto \{1, 2, 3, 4\}, \nonumber \\
   \label{eq:pidefinition}
\end{align}
where 
\begin{equation}
  \rho^{A_1,B_1}_k= \frac{\mathrm{Tr}_{A_2,B_2}\{P_k \varrho' P_k\}}{\mathrm{Tr}\{P_k \varrho'\}}.
\end{equation}
This function can be arbitrarily modified to suit different implementations. A possibility is to choose a greedy policy:

\begin{align}
\label{eq:piargmax}
   \pi= \underset{k=1,2,3,4}{\operatorname{argmax}} \left[\mathcal{C}(\rho^{A_1,B_1}_k)\right].
\end{align}

In this case, one measures the subsystem $(A_2, B_2)$ along all four measurement directions and keeps the measurement that corresponds to the highest concurrence. Alternatively, we also consider the case with only one of the measurement results, e.g., $k=1$ ($m=n=0$). We will explore both these policies in our optimization.

\end{enumerate} 

If we employ $\Pi=V(\boldsymbol \alpha_{AB})$ as the entangling operation, the output density matrix $\rho_{\text{out}}$ of the variational purification protocol will depend on the value of  $\boldsymbol \alpha_{AB}$:
\begin{equation}
\rho \to \rho_{\text{out}} (\boldsymbol \alpha_{AB}). 
\label{eq:rho_out}   
\end{equation}

\section{Numerical methods}
\label{sec:III}

\subsection{Markov chain Monte Carlo sampler}
\label{sec:IIIA}

In this approach, the main mathematical object is the 
vector space $M_4(\mathbb{C})$ of $4 \times 4$ matrices with complex entries. This vector space with the Hilbert-Schmidt inner product $\langle A, B\rangle_{HS}=\mathrm{Tr}\{A^\dagger B\}$ where $A,B \in M_4(\mathbb{C})$ is a $16$-dimensional Hilbert space. In this vector space, self-adjoint matrices form a subspace. This subspace can be identified with the Euclidean space $\mathbb{R}^{16}$, if the orthonormal basis is constructed from tensor products of Pauli matrices and the $2 \times 2$ unit matrix $I_2$. There are also other possibilities, like the Gell-Mann-type basis of $SU(4)$ or the Weyl operator basis \cite{Bertlmann}, and they either result in the same Euclidean structure or are unsuitable for our numerical approach based on $\mathbb{R}^{16}$. If $\rho$ is a density matrix, then
\begin{equation}\label{eq:convex_body}
\rho=\sum^{16}_{i=1} a_i B_i \quad \text{with} \quad 
\mathrm{Tr}\{\rho\}=1, \quad \rho \geqslant 0,  
\end{equation}
where the $B_i$ are the basis vectors and $a_i \in \mathbb{R}$ for all $i$. The positive semidefinite condition $\rho \geqslant 0$ implies that the $a_i$ have to fulfill three conditions based on Newton identities and Descartes' rule of signs \cite{Kimura, Sauer}. Finally, one arrives at the result that all $4\times 4$ density matrices are presented by a $15$ dimensional convex body $K$ around the origin of $\mathbb{R}^{15}$, because due to $\mathrm{Tr}\{\rho\}=1$ one real parameter out of sixteen is fixed. Hence, if we consider $B_{16}=I_4/2$ then $a_{16}=1/2$  and the origin of $\mathbb{R}^{15}$ is the maximally mixed state. The vector ${\bf a}=(a_1, a_2, \dots a_{15})^T \in \mathbb{R}^{15}$  with the positive semidefinite condition yields a complete description of $K$. The hit-and-run algorithm introduced by Smith \cite{Smith}
realizes a random walk inside $K$ and it was shown that the underlying Markov chain converges to the uniform stationary distribution in polynomial time \cite{Lovasz}. The convergence to the uniform distribution is independent of the starting point inside $K$ \cite{Vempala}. This algorithm provides a fast method of sampling large numbers ($10^6-10^7$) of density matrices. There exist other numerical approaches \cite{Sauer, Shang, Zyczkowski_2001, Seah, Strunz} that can also sample uniformly distributed density matrices, albeit with longer running times. All the statistical evaluations of the purification protocols are based on samples generated by the hit-and-run algorithm, whose details are shown in Appendix \ref{sec:HR}.
\begin{figure*}[ht!]
  \hspace{-0.5cm}
    \centering
    \includegraphics[width = \linewidth]{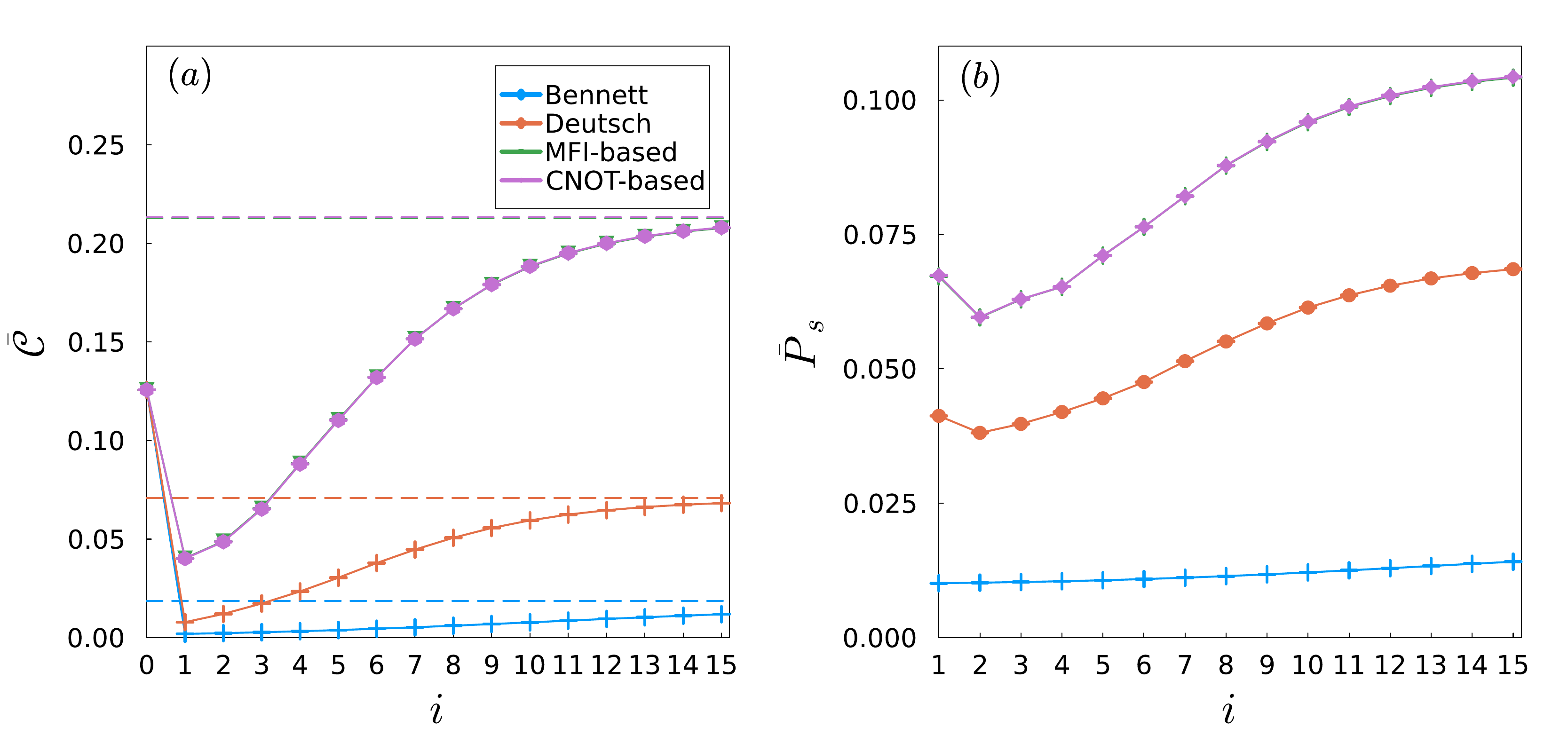}
    \caption{(a) Average concurrence and (b) success probability as a function of the number of iterations for which a sample of $N=10^7$ density matrices was used (ten runs of the hit-and-run algorithm, each one outputting $10^6$ density matrices). Numerical evaluations are done for the four different purification protocols presented in Secs. \ref{sec:IIA}--\ref{sec:IID}, i.e., Bennett, Deutsch, MFI-based, and CNOT-based protocols, respectively. Horizontal dashed lines show the theoretical limits of the protocols for the average concurrence, which are given in Eqs.~\eqref{eq:Bennettlim}--\eqref{eq:CNOTlim} for the Bennett, Deutsch, MFI-based, and CNOT-based protocols, respectively. The points are connected by lines to guide the eye. The MFI-based protocol and our proposed CNOT protocol turn out to produce the same average values, to the point that the green points are barely visible beyond the purple ones. The standard errors of the means are not visible in the plots. Values for the unbiased sample variance are available in Appendix \ref{appendix:Stat}.}
    \label{fig:protocolcomparison}
\end{figure*}
\subsection{Optimization with a quasi-Newton method}
\label{sec:IIIB}

Our task is to improve the concurrence of the output state, which is a nonlinear function of $\boldsymbol \alpha_{AB}$ (see Sec. \ref{sec:IIE}). In addition, based on Eq. \eqref{eq:anglecond} $\boldsymbol \alpha_{AB}$
has lower and upper bounds. Then, by using Eq.~\eqref{eq:rho_out} we  define the cost function $f:\mathbb{R}^{30} \rightarrow 
\mathbb{R}$ as
\begin{align}
    f(\boldsymbol \alpha_{AB}) = 1 - \mathcal{C}\left[\rho_{\text{out}}(\boldsymbol \alpha_{AB})\right]. \label{eq:f}
\end{align}

The cost function is based on the concurrence of the two-qubit state emerging after one iteration of the protocol. Let us consider that the sample of two-qubit density matrices consists of $N$ elements $\rho_1, ...,\rho_{N}$, which are mapped onto $\rho_{\text{out},1}, ...,\rho_{\text{out},N}$. The average output concurrence is estimated by
\begin{align}
    \bar{\mathcal{C}} (\boldsymbol \alpha_{AB}) =  \frac{1}{N} \sum_{j=1}^N \mathcal{C}\left[\rho_{\text{out},j}(\boldsymbol \alpha_{AB})\right].
\end{align}
Then, the average cost function reads
\begin{equation}
 \bar{f}(\boldsymbol \alpha_{AB}) =1 - \bar{\mathcal{C}}(\boldsymbol \alpha_{AB}), \label{eq:cost}
\end{equation}
which ensures that the optimal unitary transformations characterized by $\boldsymbol \alpha_{AB}$ increase the average concurrence of the whole sample. This method is the extension of the one developed in our previous investigation \cite{Preti}, where the cost function depends only on $15$-dimensional vectors, and the optimal search is constrained to specific families of two-qubit states. We implement again one of the most effective quasi-Newton methods, the L-BFGS-B optimization algorithm \cite{Nocedal,Byrd}. The gradient $\nabla \bar{f} (\boldsymbol \alpha_{AB})$ is obtained via automatic differentiation \cite{jax2018github}. The algorithm does not require second derivatives, because the Hessian matrix is approximated. This approach yields a local minimum $\boldsymbol \alpha^*_{AB}$ of $\bar{f}$, i.e., $\bar{f}(\boldsymbol \alpha^*_{AB})\leqslant \bar{f}(\boldsymbol \alpha_{AB})$ for all $\boldsymbol \alpha_{AB}$ in the Euclidean norm defined neighborhood of $\boldsymbol \alpha^*_{AB}$.
Now, we briefly describe the methodology used to optimize the variational purification protocols of Sec. \ref{sec:IIE}. The optimization is subject to general two-qubit density matrices parametrized by a $15$-dimensional real vector $\boldsymbol{a}$. As the number of samples needed to cover the space of two-qubit quantum states with sufficient precision is particularly large ($10^6-10^7$ samples; see \cite{Sauer2}), the algorithm struggles with a particularly slow optimization. Therefore, as a first attempt, we use smaller subsets of density matrices to find optimal unitary matrices and then test their performance on the whole sample.

\begin{algorithm}[H]
		\setstretch{1.20}
		\caption{Optimization with the hit-and-run (HR) given in Algorithm \ref{alg:HR} }
		\label{alg:ouralgo}
		\hspace*{\algorithmicindent} \textbf{Input} $\hbrho(\boldsymbol{a}), \boldsymbol{a} \in \mathbb{R}^{15}$,  ${P}_1,\ {P}_2,\ {P}_3,\ {P}_4$ as in Eqs.~\eqref{eq:P00},~\eqref{eq:P01}, ~\eqref{eq:P10}, and~\eqref{eq:P11}, $L_{\text{max}}$ iterations, optimizer (OPT)\\
		\hspace*{\algorithmicindent} \textbf{Output} $\hbrho(\boldsymbol{a})$
		\begin{algorithmic}[1]
        \State $\boldsymbol{a}_0 = \boldsymbol{0}$
		\For{$j=1$ to $N$}
		\State $\boldsymbol{a}_j = \text{HR}(\boldsymbol{a}_{j-1})$
		\State $\hbrho_j = \hbrho(\boldsymbol{a}_j)$
		\EndFor
			\For{$i=1$ to $L_{\text{max}}$} \Comment{with random restart}
			\State $V(\boldsymbol \alpha_{AB}) = U^{A_1, A_2}(\boldsymbol \alpha_A) \otimes U^{B_1, B_2}(\boldsymbol \alpha_B)$
			\For{$j=1$ to $N$}
                \State $\boldsymbol{\varrho}^j=\hbrho^{A_1, B_1}_j\otimes \hbrho^{A_2, B_2}_j$
                \State $\hbsigma^{j}(\boldsymbol \alpha_{AB}) = V(\boldsymbol \alpha_{AB}) \boldsymbol{\varrho}^j  V^\dagger(\boldsymbol \alpha_{AB})$
            \For{$k=1$ to $4$}
			\State $\hbrho^{jk}(\boldsymbol \alpha_{AB}) =  \frac{\mathrm{Tr}_{A_2,B_2}\{P_k \hbsigma^{j}(\boldsymbol \alpha_{AB})  P_k\}}{\mathrm{Tr}\{P_k\hbsigma^{j}(\boldsymbol \alpha_{AB})  \}}$
			\EndFor
            \State $k_{\text{max}} = \pi(\hbrho^{j1}, \hbrho^{j2}, \hbrho^{j3}, \hbrho^{j4})$
            \State $\hbrho^{j}(\boldsymbol \alpha_{AB}) = \hbrho^{j k_{\text{max}}}(\boldsymbol \alpha_{AB})$
            \EndFor
			\State $\bar{f}(\boldsymbol \alpha_{AB}) =  1 - \frac{1}{N}\sum_{j=1}^N \mathcal{C}\left[\hbrho^{j}(\boldsymbol \alpha_{AB}) \right]$
			\State $\boldsymbol \alpha_{AB}^* = \text{OPT}\left[\bar{f}(\boldsymbol \alpha_{AB}), \nabla_{\boldsymbol \alpha_{AB}} \bar{f}(\boldsymbol \alpha_{AB})\right ]$
			\For{$j=1$ to $N$}
			\State $\hbrho_j = \hbrho^{j}(\boldsymbol \alpha_{AB}^*) $
			\EndFor
			\EndFor
		\end{algorithmic}
\end{algorithm}

The general optimization routine is presented in Algorithm \ref{alg:ouralgo}. First, general density matrices are sampled, which is followed by the optimization of the average cost function. The output density matrices are reinserted in the protocol for a successive purification round. This is also optimized, until the maximal number
of iterations $L_{\text{max}}$ is reached. In every step of the iteration, we find different optimal unitary matrices, which yields an adaptive purification protocol \cite{Preti}.

\section{Results}
\label{sec:IV}

\subsection{Statistics of purification protocols}
\label{sec:IVA}

In this section, we compare the recurrence protocols
of Secs. \ref{sec:IIA}, \ref{sec:IIB}, \ref{sec:IIC}, and \ref{sec:IID} based on their average performance on a sample of two-qubit states drawn from an almost uniform distribution \cite{Sauer}. More formally, let $\rho_1, ...,\rho_{N}$ be $N$ density matrices generated by the hit-and-run algorithm. As the composition of completely positive maps is again completely positive \cite{Paulsen}, every entanglement purification protocol presented in this work acts as a completely positive trace-preserving non-linear map. We denote this by $\Phi$, which maps the set of two-qubit quantum states
\begin{align}
    D(\mathbb{C}^4) = \{\rho \in M_4(\mathbb{C}) : \rho \geq 0, \, \Tr{\rho} = 1\},
\end{align}
i.e., positive semidefinite matrices with unit trace,
onto itself
\begin{align}
    \Phi: D(\mathbb{C}^4) \mapsto D(\mathbb{C}^4), \quad \Phi(\rho) = \rho'.
\end{align}

A purification map $\Phi$ is iteratively applied to a density matrix to purify it towards a maximally entangled state, that is, to extract a state with higher concurrence. If successful, the protocol approaches usually the concurrence value $\mathcal{C}=1$ in the limit of infinite iterations, but there are also known cases of one-step purifiable states \cite{Bennett2, Torres}. If unsuccessful, then a state with non-zero concurrence is mapped to a state with zero concurrence, thereby destroying entanglement, which happens at the first iteration of each protocol. The average concurrence after $i$ iterations is estimated by
\begin{align}
\label{eq:concav}
    \bar{\mathcal{C}}^{(i)}  =  
    \frac{1}{N} \sum_{j=1}^N \mathcal{C}\left[\Phi^{i}(\rho_j)\right].
\end{align} The sample standard deviation reads
\begin{align}
\label{eq:concsstdev}
    s^{(i)}_\mathcal{C}=  
    \sqrt{\frac{1}{N-1} \sum_{j=1}^N \left\{\mathcal{C}\left[\Phi^{i}(\rho_j)\right] - \bar{\mathcal{C}}^{(i)}\right\}^2}
\end{align}
and the standard error, i.e., the standard deviation of $\bar{\mathcal{C}}^{(i)}$, is given by:
\begin{equation}\label{eq:concerror}
\sigma^{(i)}_{\bar{\mathcal{C}}}=\frac{s^{(i)}_\mathcal{C}}{\sqrt{N}}.
\end{equation}
Similarly, the average success probability can be estimated. In this case, we should first point out that for a general two-qubit density matrix $\rho$ the success probability can only be defined for those states whose concurrence is non-zero, that is in those cases where the entanglement purification protocols are not failing. The probability of a measurement at the $i$-th iteration can be described as
\begin{align}\label{eq:success_prob}
    &p_{k}=\Tr{ P_{k} \Phi^{i}(\rho_j)}, \quad k\in\{1,2,3,4\},\\
    &P^{(i)}_s(k, \rho_j) = \begin{dcases}
    p_{k} & \text{if } C\left[\Phi^{i}(\rho_j)\right] \geq 0, \\
    0              & \text{otherwise},
    \end{dcases} \label{eq:concurrencelzero}
\end{align}
where $P_{k}$ is the projector on $(A_2, B_2)$ defined in Eqs. \eqref{eq:P00}, \eqref{eq:P01}, \eqref{eq:P10}, and \eqref{eq:P11}. In the case of the protocols presented in Secs. \ref{sec:IIA}, \ref{sec:IIB}, \ref{sec:IIC}, different measurement results yield the same probability, and analytical formulas are given for the success probability $P^{(i)}_s(\rho_j)$. In the case of the MFI-based protocol, the success probability is given by the probability of successfully performing the quantum operation given in Eq.~\eqref{eq:M} and the probability of measuring the state in one of the four possible outcomes is $1/4$. The success probability of the approach in Sec. \ref{sec:IID} is also shown. However, in the optimized variational case, $P^{(i)}_s(\rho_j)$ is a function of $P^{(i)}_s(k, \rho_j)$, where $k$ is selected by the policy $\pi$ introduced in Eq.~\eqref{eq:pidefinition}. It is worth noting that the condition in Eq. ~\eqref{eq:concurrencelzero} is decisive only in the first step for the protocols in Secs. \ref{sec:IIA} and \ref{sec:IIB}, because after that all the remaining entangled two-qubit states are improved towards a Bell state. The average success probability is estimated after each iteration step by 
 \begin{align}
 \label{eq:probav}
    \bar{P}^{(i)}_s = \frac{1}{N} \sum_{j=1}^N P^{(i)}_s(\rho_j),
\end{align}
which corresponds to the probability of successfully implementing the $i$-th step of the entanglement purification protocol on an unknown two-qubit state with non-zero concurrence. In this case, the standard deviation of the sample is then given by
\begin{align}
\label{eq:probsstdev}
    s^{(i)}_{P_s}=  
    \sqrt{\frac{1}{N-1} \sum_{j=1}^N \left\{P^{(i)}_s(\rho_j) - \bar{P}^{(i)}_s\right\}^2}
\end{align}
with the standard error
\begin{equation}\label{eq:proberror}
\sigma^{(i)}_{\bar{P}_s}=\frac{s^{(i)}_{P_s}}{\sqrt{N}}.
\end{equation}

\begin{figure*}
  \hspace{-1cm}
    \centering
    \includegraphics[width = \linewidth]{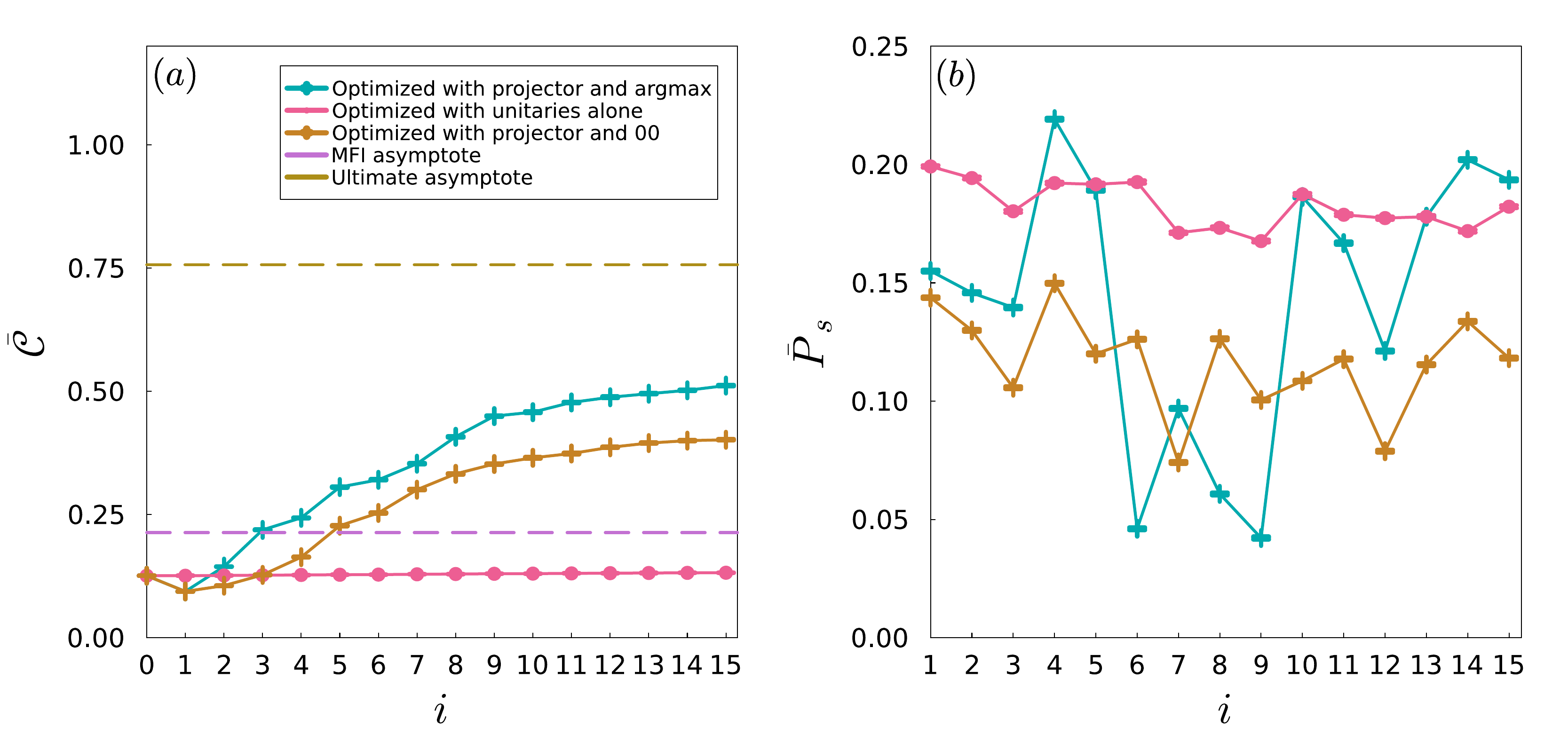}
    \caption{(a) Average concurrence and  (b) success probability of the optimized purification protocols as a function of the number of iterations. The values shown in the plot represent the average over $N=10^7$ density matrices (ten runs of the hit-and-run algorithm, each one outputting $10^6$ density matrices). The asymptote of the MFI-based protocol in Eq. \eqref{eq:MFIlim} and the ultimate limit of Eq. \eqref{eq:Ulitmatelim} are also shown. The points are connected by lines to guide the eye. The strategy of first destroying entanglement and then improving it (green points) delivers the best performance compared also to the protocols of Fig. \ref{fig:protocolcomparison}. If we only keep the measurement result $k = 1$ $(m=0, n=0)$, the average concurrence turns out to be lower compared to the case where the argmax policy of Eq.~\eqref{eq:piargmax} is used. The standard error of the mean is not visible in the plot. Values for the unbiased sample variance are available in Appendix \ref{appendix:Stat}.}
    \label{fig:opt_concurrence}
\end{figure*}

The initial average concurrence is calculated over $N=10^7$ density matrices and yields
\begin{equation}
 \bar{\mathcal{C}}^{(0)}=0.1257(2), \label{eq:C0}
\end{equation}
which represents the amount of entanglement present in the sample. Here we use parentheses to denote the standard error. We consider this number as a threshold and expect that entanglement purification protocols can improve it. The results of the four protocols are shown in Fig.~\ref{fig:protocolcomparison}. It is immediate to see that all protocols destroy entanglement in their very first iteration. The twirling operations used in both the Bennett and Deutsch protocols to obtain Werner or Bell diagonal
states, respectively, destroy a significant portion of the entangled states. The remaining entangled states reach $\mathcal{C}=1$ in the limit of infinite iterations. We find these limits by imposing the conditions given by Eqs.~\eqref{eq:Bennettcond} and \eqref{eq:Deutschcond} on the whole sample. If a state does not satisfy it, then we assign $\mathcal{C}=0$ to it, otherwise, we set $\mathcal{C}=1$. In other words, every state satisfying the conditions can be purified into a Bell state in the limit of infinite iterations. This limit for the Bennett protocol yields
\begin{equation}
\label{eq:Bennettlim}
 \bar{\mathcal{C}}^{(\infty)}_{\text{Bennett}}=0.01865(4),
\end{equation}
while the Deutsch approach results in
\begin{equation}
\label{eq:Deutschlim}
 \bar{\mathcal{C}}^{(\infty)}_{\text{Deutsch}}=0.0709(1).
\end{equation}
We would like to stress that this value is lower than the initial average concurrence in Eq.~\eqref{eq:C0} for both protocols. Furthermore, they require several iterations to approach these limits. 
The other two protocols yield 
\begin{equation}
\label{eq:MFIlim}
 \bar{\mathcal{C}}^{(\infty)}_{\text{MFI}}=0.2128(1),
\end{equation}
and
\begin{equation}
\label{eq:CNOTlim}
 \bar{\mathcal{C}}^{(\infty)}_{\text{CNOT}}=0.2133(1),
\end{equation}
which are based on the conditions in Eqs.~\eqref{eq:MFIcond1}, \eqref{eq:MFIcond2}, \eqref{eq:CNOTcond1}, and \eqref{eq:CNOTcond2}. The performance of these protocols is almost the same and despite destroying entanglement in the first iteration, they can improve the average concurrence of the sample beyond its starting value. They achieve this already after four iterations (see Fig.~\ref{fig:protocolcomparison}). It is worth noting again that the asymptote of the Bennett protocol contains only the Bell state $\ket{1}$, being a stable fixed point, while the other three protocols 
have two relevant stable fixed points in the limit of infinite iterations.
These findings together with the average success probabilities characterize the performances and they show that both the MFI-based and our CNOT protocols are superior to the pioneering ones. However, one can define an ultimate limit of every known or future proposal for entanglement purification, namely, when all entangled two-qubit states are converted to a maximally entangled state, which for our sample results in the numerical estimate
\begin{equation}
 \bar{\mathcal{C}}^{(\infty)}_{\text{ultimate}}=0.7569(1).
 \label{eq:Ulitmatelim}
\end{equation}
Given this number, one can conclude that all four evaluated protocols are not very effective. The performance is confirmed by the analysis of the fidelities with respect to the stable fixed-point states of the protocols, i.e., $r_1$ for the Bennett, $r_4$ and $r_2$ for the Deutsch, $r_1$ and $r_2$ for the MFI and $r_4$ and $r_2$ for the CNOT protocols, as a function of the number of protocol iterations, which are shown in Appendix \ref{appendix:sef}. This raises the question of how one can obtain better performance, which will be discussed in the subsequent section.

\subsection{Optimization of variational recurrence protocols}
\label{sec:IVB}

In this section, we investigate numerically the variational purification protocol described in Sec. \ref{sec:II}. The optimization is based on the method given in Sec. \ref{sec:IIIB}, where the gradient of the cost function associated with the protocol is computed through automatic differentiation. First, we generate $N=10^6$ density matrices using the hit-and-run algorithm to obtain an almost uniformly distributed sample. Afterward, to speed up the optimization, we randomly pick $N_s = 1000$ density matrices from this sample and calculate the gradient. This approach guarantees that optimization is less time consuming and the convex set of all two-qubit states is well represented. Nevertheless, the sampling of only $10^3$ density matrices is not enough to ensure uniform distribution, in particular for the statistics of the CNOT and MFI protocols. The optimization results in a numerical value for $\boldsymbol \alpha^*_{AB}$. We then apply the $\boldsymbol \alpha^*_{AB}$-dependent protocol on the whole sample. We also try different strategies for the measurement policy $\pi$ of Eq.~\eqref{eq:pidefinition}. In particular, we test the policy of Eq.~\eqref{eq:piargmax} together with the case in which we only track the measurement result $k=1$ ($m=n=0$) in each iteration. 

Our first try is the case when $\Pi$ in Eq. \eqref{eq:varprot} is equal to $V(\boldsymbol \alpha_{AB})$ with policy $\pi$ of Eq. \eqref{eq:piargmax}. In Fig. \ref{fig:opt_concurrence} we present the result, which demonstrates that this strategy does not destroy initial entanglement, as one expects. However, the actual increase in average concurrence is quite limited, which implies that an enormous number of iterations are needed to purify the density matrices towards a maximally entangled state. As we seek to obtain a more efficient protocol, we need to consider additional operations performed on the sample of density matrices.

An interesting feature of Sec. \ref{sec:IVA} is that every protocol first destroys entanglement before it starts to improve the remaining two-qubit quantum states. Next, we consider the hypothesis that entanglement must be destroyed in order to find an efficient protocol. Therefore, for the first iteration, we use the projector
\begin{eqnarray}
\Pi=M^{A_1,A_2}_2 \otimes  M^{B_1,B_2}_2 \quad \text{with} \quad M_2 = I_4 - \ket{2}\bra{2} \nonumber \\
\label{eq:special_projector}
\end{eqnarray}
as the operation of the purification protocol, and for the rest of the iterations we make use of the optimization with $\Pi=V(\boldsymbol \alpha_{AB})$, which leads to higher average concurrence. Here, we have tested both of our measurement policies $\pi$. In both cases, these approaches outperform the limits of the MFI-based and the CNOT protocols, which are shown in Fig.~\ref{fig:opt_concurrence}. We remind the reader that every optimized step of the variational protocol yields a different entangling gate. The resulting performance for many iterations seemingly approaches a limit, which is almost halfway between the ultimate asymptote and the asymptote defined by the MFI-based or the CNOT protocol. The average success probability of this optimized protocol oscillates as a function of the number of iterations. We do not have a proper explanation for this effect, as it depends on the non-linear optimizer. However, we can see that the values of the success probability are almost always higher than those produced by the protocols considered in Fig.~\ref{fig:protocolcomparison}.

Finally, we need to address the realization of these optimized and abstract protocols. Concerning the two-qubit gates in $V(\boldsymbol \alpha_{AB})$, one can always employ quantum compilation strategies \cite{Martinez, Debnath, osti_1785933, preti2024hybrid}, where the optimal unitary matrix $V(\boldsymbol \alpha^*_{AB})$ is translated into native gates on the chosen experimental platform. However, we are not aware of any possible implementation of $M_2$ in Eq.~\eqref{eq:special_projector}, but it seems necessary to first destroy entanglement before any variational purification protocol is applied. 


\section{Summary and conclusions}
\label{sec:V}

To summarize, we have presented a numerical method that is capable of characterizing the performance of entanglement purification protocols. We have presented a CNOT-based protocol, which was evaluated together with two pioneering protocols and a recent proposal based on matter-field interactions. Our results show that all the protocols destroy entanglement in the very first iteration. This was known for the pioneering protocols, and here in addition we have demonstrated quantitatively that only a small set of entangled states is kept. Even though these states are turned into a Bell state in the limit of infinite iterations, the average concurrence of the whole sample stays below its starting value. The MFI-based and the CNOT-based protocols perform better and they can turn slightly more than $21 \%$ of the two-qubit states into a Bell state.

We have defined the ultimate limit of all possible purification protocols, which is nothing other than the percentage of all entangled states within the set of all quantum states, i.e., approximately $75\%$. In other words, an ultimate protocol can purify all entangled states into maximally entangled ones. In this context, we have searched for optimal asymmetric entangling gates in the nodes $A$ and $B$. We have found that this approach is improving the average concurrence very slowly. Therefore, motivated by the other approaches, we have included in the first iteration an entangling projection, which destroys some entanglement. This strategy turns out to be a boost for the optimized variational approach, which can outperform all protocols discussed in this paper. However, even the variational approach seems unable to reach the ultimate bound for entanglement purification. At the current stage, we cannot determine whether this result could be improved by implementing different quantum operations or optimization methods. There might also be an upper bound for this family of protocols that lies below the ultimate asymptote.

Our numerical analysis focuses on the improvement of the average concurrence. However, we would like to point out that all the protocols investigated in this work have a common property. If some information about the input state is known, e.g., the state has an overlap strictly larger than $0.5$ with one of the Bell states, then we know beforehand whether or not these protocols convert entangled states into a fixed maximally entangled state, which is usually a Bell state. A further difficulty is that the iterations of the protocol are nonlinear quantum state transformations, which lead to chaotic behavior \cite{Kiss2011, Guan2013, Gilyen2016, Portik2022}. Therefore, in the trace norm topology \cite{Paulsen}, seemingly close density matrices might have different future trajectories. Our approach avoids this interesting but complicated behavior by using the average concurrence, a choice, which we have demonstrated to be also successful in finding different and effective entanglement purification protocols.

Finally, some comments on the average success probabilities are in order. It is known that improving the concurrence alone is not a good enough measure, because success probabilities play a crucial role in the identification of required resources, i.e., how many qubits are required to perform some iterations. To improve the success probabilities as well, this leads to a multiobjective optimization task. This is not included here, because this work aims to introduce a general evaluation scheme based on the hit-and-run algorithm and the concurrence, which have given insight into the performance of entanglement purification protocols and shown that improvements are possible in computer-based protocol designs.  

\begin{acknowledgments}
This work was supported by AIDAS-AI, Data Analytics
and Scalable Simulation, which is a Joint Virtual Laboratory gathering the Forschungszentrum J\"ulich and the French Alternative Energies and Atomic Energy Commission, by the Deutsche Forschungsgemeinschaft (DFG, German Research Foundation) under Germany’s Excellence Strategy – Cluster of Excellence Matter and Light for Quantum Computing (ML4Q) EXC 2004/1 – 390534, and by the Jülich Supercomputing Center. We are grateful to J. M. Torres and T. Calarco for stimulating discussions. Simulations were realized in PYTHON using the library JAX \cite{jax2018github} and in JULIA. The code and the data are available at \href{github.com/franz3105/soepp}{\url{https://github.com/franz3105/soepp}} (code) and \href{10.5281/zenodo.10996551}{\url{https://zenodo.org/uploads/10996551}} (data sets).

\end{acknowledgments}

\appendix

\section{Hit-and-run algorithm}
\label{sec:HR}

In this Appendix we briefly describe the steps of the hit-and-run algorithm in Algorithm \ref{alg:HR}. Given a $K \subseteq \mathbb{R}^{15}$, we generate for an $ {\bf a} \in K$ a random uniform vector 
${\bf x}$ on the sphere, which is around $ {\bf a}$ and has unit radius. We generate a random uniform number $\lambda$ on the interval $[-\sqrt{3}/2,\sqrt{3}/2]$, because $K$ is inside the sphere of radius $\sqrt{3}/2$ around the origin \cite{Sauer}. If ${\bf a}'={\bf a}+\lambda {\bf x} \in K$, then we move there, otherwise we start all over from 
${\bf a}$. We always start the sampling from ${\bf a}={\bf 0}$, i.e., the maximally mixed state.

\begin{algorithm}[H]
		\setstretch{1.20}
		\caption{Hit-and-run}
		\label{alg:HR}
	\begin{algorithmic}[1]
		\State $j=1$ and 
		${\bf a}^{(1)}={\bf 0}$.
	\While {$j < N$}
		\State ${\bf x}^{(j)} \sim \mathcal{N}({\bf 0},I_{15})$.
		\State ${\bf x}^{(j)} = {\bf x}^{(j)}/ \| {\bf x}^{(j)}\|$
		\State Set $I=[-r,r]$  with $r=\sqrt{3}/2$.
		\State $m=0$
	\While {$m = 0$}	
		\State $\lambda \sim \mathcal{U}_I$. \label{marker}
	 \If{${\bf a}^{(j)}+\lambda {\bf x}^{(j)} \in K$}
       \State ${\bf a}^{(j+1)}={\bf a}^{(j)}+\lambda {\bf x}^{(j)}$
       \State  $j=j+1$
       \State $m=1$
     \Else
      \If {$\lambda>0$}
          \State $I=[-r,\lambda]$
        \Else
          \State $I=[\lambda,r]$
      \EndIf
     \EndIf
     \EndWhile
	\EndWhile
	\end{algorithmic}
\end{algorithm}

\begin{figure*}
    \centering
    \hspace{-0.4cm}
    \includegraphics[width = \linewidth]{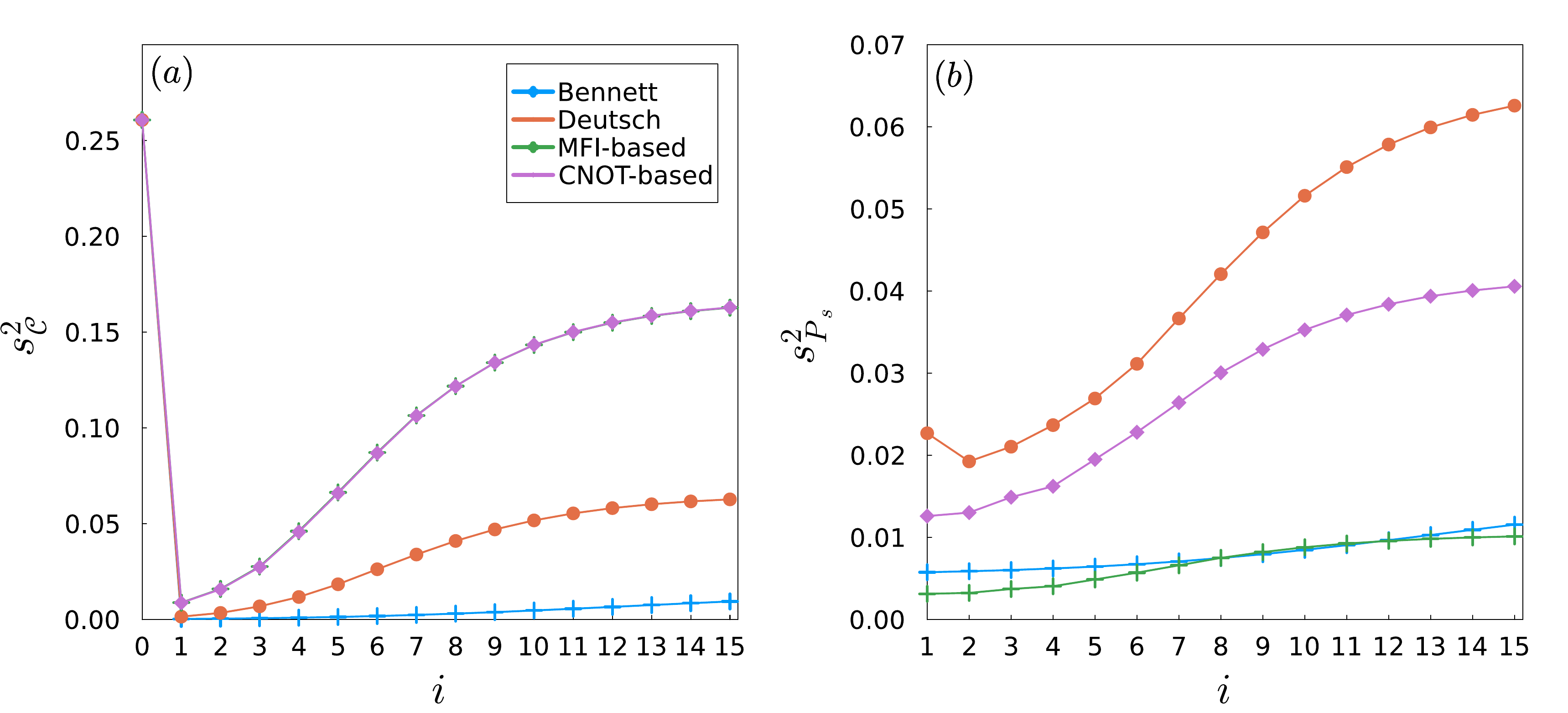}
    \includegraphics[width = 0.98\linewidth]{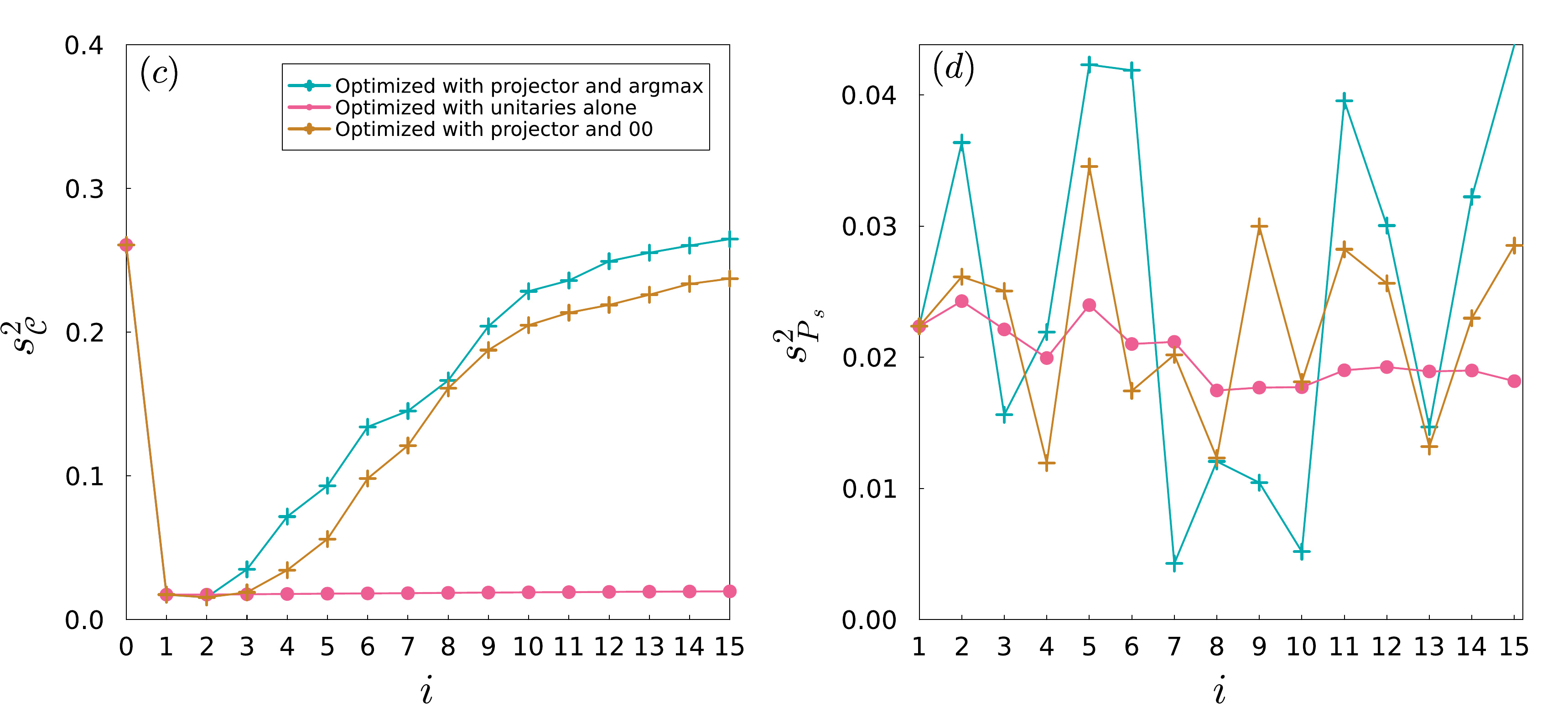}
    \caption{Unbiased sample variances of the concurrence and the success probability as a function of the number of iterations for  (a) and (b) the non-variational and (c) and (d) the optimized variational protocols. We use the same color scheme as in Figs. \ref{fig:protocolcomparison} and \ref{fig:opt_concurrence}. We observe that the sample variance grows with the number of iterations and reaches its maximum asymptotic values when all the density matrices are all mapped to either one or zero concurrence [see Eq.~\eqref{eq:samplevar_asymptotic}]. The fraction of density matrices with non-zero concurrence in the limit of infinite iterations determines the performance of the protocol and also the maximum of the sample variance.}
    \label{fig:protocolcomparisonv}
\end{figure*}

\section{Statistics of entanglement purification}\label{appendix:Stat}

\begin{figure*}
    \centering
    \hspace{-0.5cm}
    \includegraphics[width=\linewidth]{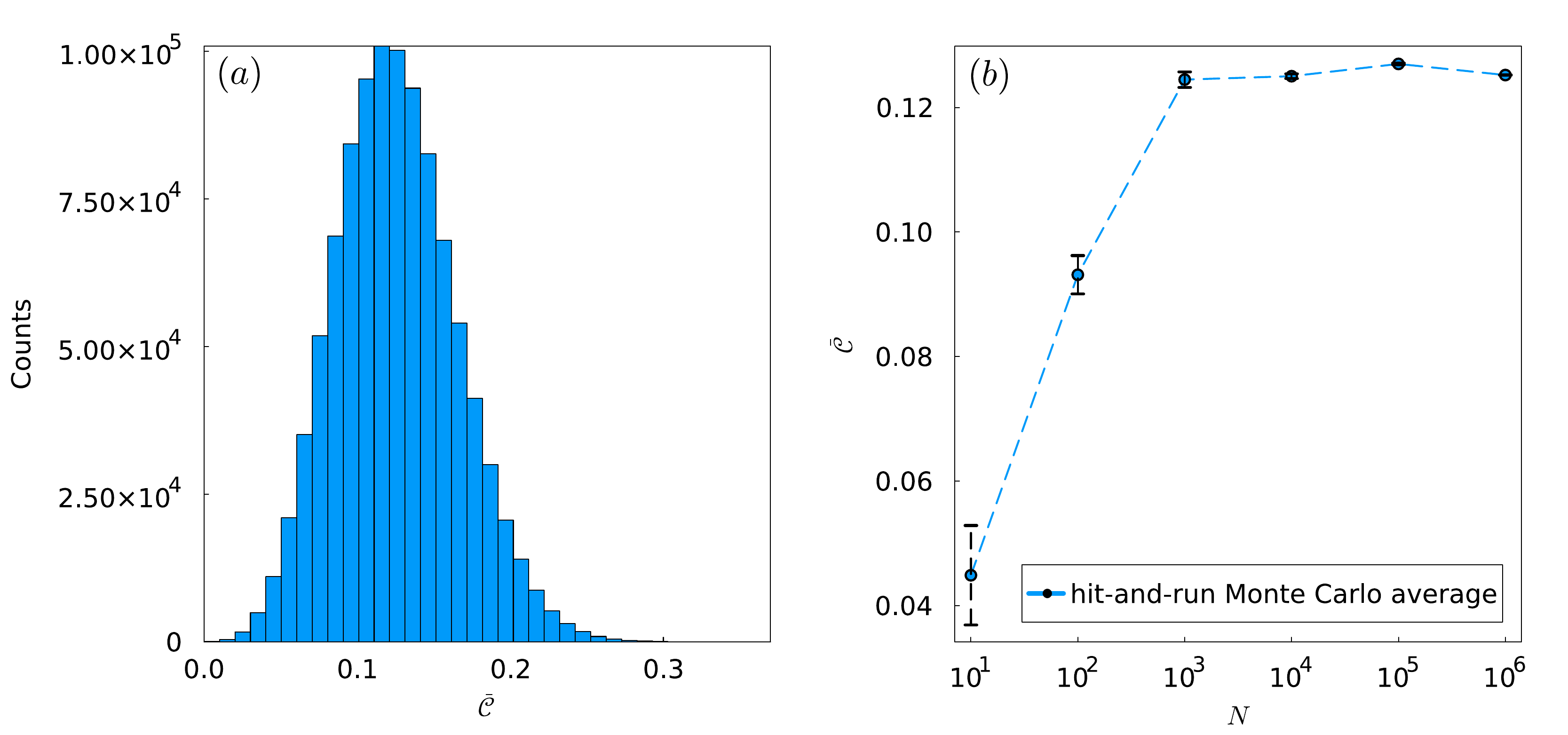}
    \caption{(a) Bar chart of the distribution of the concurrence for a sample of two-qubit density matrices after removing the states with concurrence zero; and (b) average initial concurrence as a function of the number of samples generated by the hit-and-run Monte Carlo algorithm. Vertical bars show the standard error of the mean. We see that the average approximately stabilizes after $N=10^3$ samples. This is only true for the initial concurrence, as the average concurrence after each run of the purification protocol needs a higher resolution (empirically we find $N > 10^4$, especially for the MFI protocol). Plot (a) uses $10^6$ density matrices, obtained by running the hit-and-run algorithm once.}
    \label{fig:inital_concurrence_dist}
\end{figure*}

In this Appendix, details concerning the sample mean and standard deviation are discussed. The primary task is to estimate numerically different averages of the concurrence or the success probability over all two-qubit density matrices. In the main text we have already identified this set with the convex body $K$ in the Euclidean space $\mathbb{R}^{15}$. Thus, every density matrix $\rho$ in Eq.~\eqref{eq:convex_body} is uniquely described by a vector ${\bf a} \in K$. In this context, the average concurrence reads
\begin{align}
    \bar{\mathcal{C}}  =\frac{1}{\mbox{vol}(K)} \int_{{\bf a} \in K} \mathcal{C}\left[\rho({\bf a})\right]\, d^{15}{\bf a}
\end{align}
with respect to the Lebesgue measure in $\mathbb{R}^{15}$. If
$\rho_1, ..., \rho_N$ are generated by the hit-and-run algorithm, then the estimated value of $\bar{\mathcal{C}}$  reads $\sum_{j=1}^N \mathcal{C}(\rho_j)/N$. Similarly, the average success probability
\begin{equation}
 \bar{P}_s  =\frac{1}{\mbox{vol}(K)} \int_{{\bf a} \in K} P_s\left[\rho({\bf a})\right]\, d^{15}{\bf a}    
\end{equation}
is estimated by $\sum_{j=1}^N P_s(\rho_j)/N$. The standard deviations of these means are given in Eqs.~\eqref{eq:concerror} and \eqref{eq:proberror}. In the limit of infinite iterations, the distribution of the concurrence assumes a specific form: The entanglement purification protocol has purified a certain number $S$ of density matrices, while the other $N-S$ have concurrence zero. This means that the average concurrence in Eq.~\eqref{eq:concav} is given by
\begin{align}
\label{eq:B3}
    \lim_{i \rightarrow \infty} \bar{\mathcal{C}}^{(i)}  = \frac{S}{N},
\end{align}
and the sample standard deviation in Eq.~\eqref{eq:concsstdev} reads
\begin{eqnarray}
\label{eq:samplevar_asymptotic}
    \lim_{i \rightarrow \infty} s^{(i)}_\mathcal{C}&=&  
    \sqrt{\frac{S}{N-1} \left(1- \frac{S}{N}\right)^2+\frac{N-S}{N-1} \frac{S^2}{N^2}} \nonumber \\
    &=&\sqrt{\frac{S}{N-1}\left(1- \frac{S}{N}\right)}.
\end{eqnarray}
Now, let us assume that after the $i$th iteration quantum state $\rho_j$ belonging to the set of purifiable density matrices has the concurrence
\begin{equation}
\mathcal{C}^{(i)}(\rho_j)=1-\epsilon^{(i)}_j \quad \text{with} \quad 1>\epsilon^{(i)}_j>0.
\end{equation}
This model describes how far the concurrence of $\rho_j$ from $\mathcal{C}=1$ is, which is attained for $i \rightarrow \infty$. Now, we have
\begin{equation}
\bar{\mathcal{C}}^{(i)}  =\frac{S}{N}-\frac{1}{N}\sum^S_{j=1} \epsilon^{(i)}_j <\lim_{i \rightarrow \infty} \bar{\mathcal{C}}^{(i)},
\end{equation}
i.e., the average concurrence is always smaller than 
$S/N$ [see Eq. \eqref{eq:B3}]. Then the sample standard deviation reads
\begin{eqnarray}
\label{eq:B1}
 &&s^{(i)}_\mathcal{C}= \sqrt{\frac{1}{N-1} \sum^S_{j=1}\left(1-\epsilon^{(i)}_j- \bar{\mathcal{C}}^{(i)}\right)^2+\frac{N-S}{N-1} \left( \bar{\mathcal{C}}^{(i)}\right)^2} \nonumber \\
    &&=\sqrt{\frac{S-2\epsilon_i}{N-1}\left(1- \frac{S}{N}\right)-\frac{\epsilon^2_i}{N(N-1)}+\frac{\beta_i}{N-1}}, 
\end{eqnarray}
where
\begin{equation}
  \epsilon_i= \sum^S_{j=1} \epsilon^{(i)}_j, \quad \beta_i=  \sum^S_{j=1} \left(\epsilon^{(i)}_j\right)^2.  
\end{equation}
Since $\left(\epsilon^{(i)}_j\right)^2<\epsilon^{(i)}_j$ for all $j$, we get $\beta_i<\epsilon_i$. Hence
\begin{equation}
\label{eq:B2}
-2\epsilon_i \left(1- \frac{S}{N}\right)- \frac{\epsilon^2_i}{N}+\beta_i<-\epsilon_i \left(1- \frac{2S}{N}\right)- \frac{\epsilon^2_i}{N}.  
\end{equation}
If $2S<N$, which is the case of the protocols discussed in \ref{sec:IVA}, we obtain based on Eqs. \eqref{eq:B1} and
\eqref{eq:B2} that
\begin{equation}
 s^{(i)}_\mathcal{C} < \lim_{i \rightarrow \infty} s^{(i)}_\mathcal{C},
\end{equation}
i.e., the sample standard deviation reaches its maximum in the limit of infinite iterations. This property is shown in Fig. \ref{fig:protocolcomparisonv}.
Finally, we also want to visualize the distribution of the concurrence for the samples of density matrices that are generated by the hit-and-run algorithm using bar charts. We know that approximately $24.24$ \% of them  are separable quantum states and have concurrence zero. These cannot be used by the purification protocols discussed in the main text. Therefore, we remove the bar chart corresponding to these matrices from our histograms, as they would be simply represented by a single enormous peak around zero. The distribution of the concurrence for the remaining randomly sampled two-qubit density matrices is given in Fig.~\ref{fig:inital_concurrence_dist}. It seems that the concurrence resembles a (skew) Gaussian distribution, however, this is only a hypothesis and one should prove or disprove it by using methods developed in random matrix theory concerning density matrices \cite{yczkowski_2011}.

\begin{figure*}
    \hspace{-0.5cm}
    \centering
    \includegraphics[width=19cm]{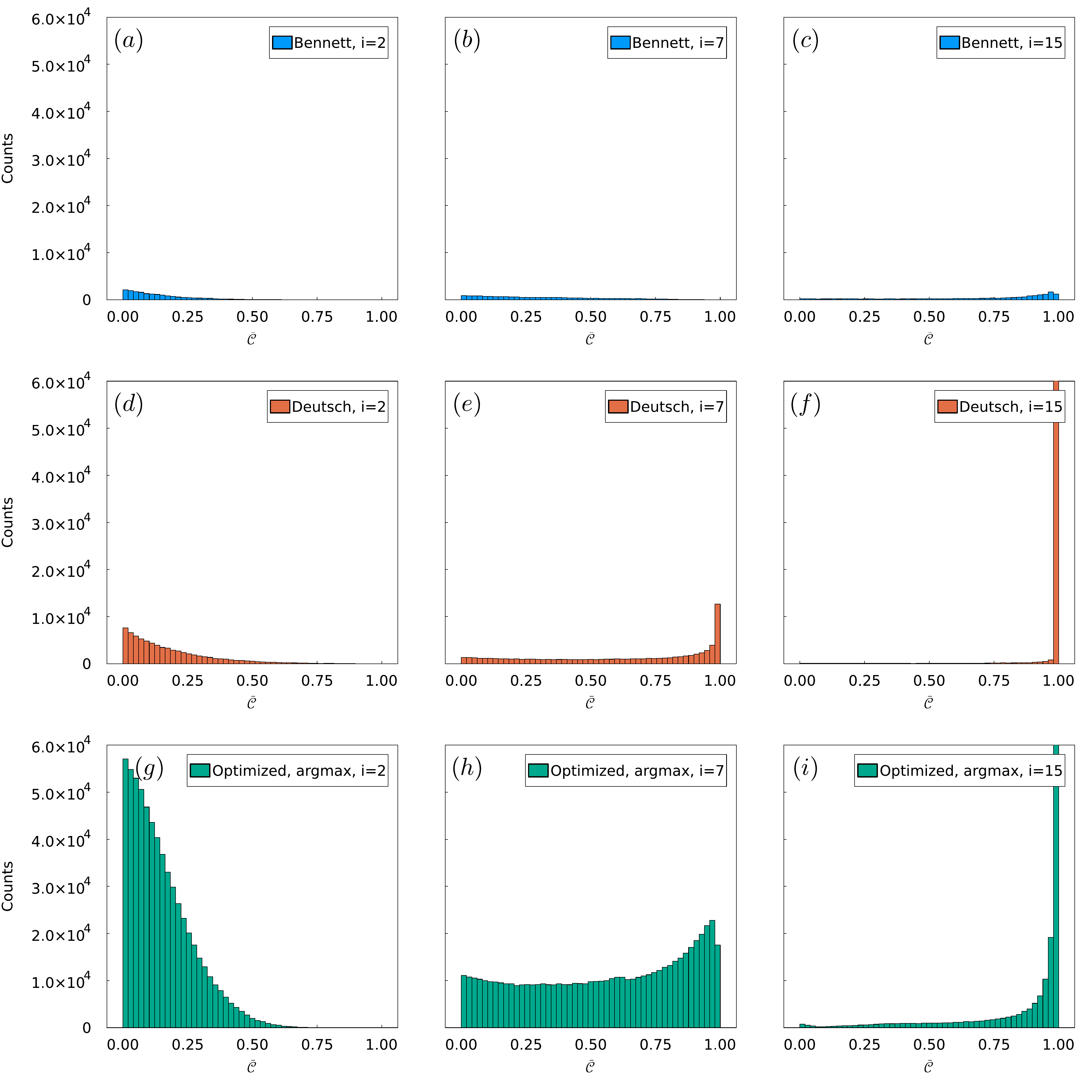}
    \caption{Bar charts representing 50 bins of the concurrence distribution for the Bennett -- (a), (b), (c), Deutsch -- (d), (e), (f), and optimized protocol [using the argmax policy in Eq.~\eqref{eq:piargmax} and the projector $\Pi$ in Eq.~\eqref{eq:special_projector}] -- (g), (h), (i), each one for the second, seventh, and fifteenth iteration -- corresponding to the first, second and third column of the plots. States with concurrence zero have been removed to allow us to visualize the action of the protocols, as they skew the distribution due to their generally high number. As a consequence, the bar charts exhibit different heights, since the three protocols map different numbers of states to concurrence zero. Vertical axes have been set so to range between 0 and $6 \times 10^4$, so that the reader can visualize the growth of the number of states with concurrence equal to one on the right as the number of iterations grows. We see that the optimized protocol preserves the most states, whereas the number of purified states for the Deutsch and Bennett protocols is significantly lower. These bar charts use $N=10^6$ samples of the concurrence (one run of the hit-and-run algorithm).}
    \label{fig:concurrence_histograms}
\end{figure*}

In Fig.~\ref{fig:concurrence_histograms}, we see the iteration-based evolution of the distribution of the concurrence for $\mathcal{C} > 0$. Each column represents one of three different points in the purification protocols: (a), (d), (g) $i=2$; (b), (e), (h) $i=7$ and (c), (f), (i) $i=15$. We show here nine plots, where the first row (a), (b), (c) represents the Bennett protocol, the second row (d), (e), (f) the MFI protocol and the third one (g), (h), (i) the CNOT protocol. We find, as expected, that the number of states with non-zero concurrence is much lower in the case of the Bennett protocol than for the other two, although every protocol improves the distribution of the concurrence towards $\bar{\mathcal{C}}=1$.

\section{Statistical evaluation of the fidelities}
\label{appendix:sef}

To provide the reader with a better understanding of how the protocols transform noisy entangled states into Bell states, we investigate here the evolution of the fidelities with respect to the stable fixed points. The study of fidelity is always conditioned on the properties of the input states. As we have already discussed in Sec. \ref{sec:II}, these properties define different stable fixed points towards which the state is mapped by the protocol. Thus, the whole sample of input states will be separated into sets according to their stable fixed points and fidelities will be evaluated only in the corresponding set. This consideration allows us to exclude situations when the output state is separable and converges towards the maximally mixed state, which has an overlap of $0.25$ with any Bell states and therefore would contribute to and at the same time skew the average output fidelity of the protocol. It is obvious that the set with the maximally mixed state as a fixed point will be not considered. This step is not necessary when one uses the concurrence. 

For a density matrix $\rho$, we consider the overlaps

\begin{align}
    r_k = \tr{\rho \ket{k}\bra{k}},
\end{align}
where the states $k=1,2,3,4$ are the Bell states defined before Eq.~\eqref{eq:bellstates}.
We then use them to define the output fidelity for the Bennett protocol \cite{Bennett1993}

\begin{align}\label{eq:fbennett}
    F_{\text{Bennett}}(\rho) =  \begin{cases}
      r_1 & \text{if $2r_1 > 1,$} \\
      0 & \text{otherwise}.
    \end{cases} 
\end{align}

Similarly, we define the output fidelities for the other protocols using their respective conditions for purification. For the Deutsch protocol, we have

\begin{align}\label{eq:fdeutsch1}
    F^{(4)}_{\text{Deutsch}}(\rho) = \begin{cases}
        r_4 & \text{if $(2r_1 - 1)(1 - 2 r_4) > 0$}, \\
        0 & \text{otherwise}.
    \end{cases}
\end{align}

\begin{align}\label{eq:fdeutsch2}
    F^{(2)}_{\text{Deutsch}}(\rho) = \begin{cases}
        r_2 & \text{if $(2r_2 - 1)(1- 2 r_3) > 0$}, \\
        0 & \text{otherwise}.
    \end{cases}
\end{align}

In the case of the MFI protocol, we get 
\begin{align}\label{eq:fmfi1}
    F^{(1)}_{\text{MFI}}(\rho) = \begin{cases}
        r_1 & \text{\footnotesize if $(2r_1-1)(1-2r_3)>-
  (2{\rm Im [r_{13}]})^2-(2{\rm Re}[r_{24}])^2$}, \\
        0 & \text{otherwise}
    \end{cases}
\end{align}

\begin{align}\label{eq:fmfi2}
    F^{(2)}_{\text{MFI}}(\rho) = \begin{cases}
        r_2 & \text{\footnotesize if $(2r_2-1)
  (1-2r_4)>-(2{\rm Im [r_{24}]})^2-(2{\rm Re}[r_{13}])^2$}, \\
        0 & \text{otherwise},
    \end{cases}
\end{align}
and finally CNOT protocol yields
\begin{align}\label{eq:fcnot1}
    F^{(4)}_{\text{CNOT}}(\rho) = \begin{cases}
        r_4 & \text{\footnotesize if $(2r_1-1)(1-2r_4)>-
  (2{\rm Im [r_{23}]})^2-(2{\rm Re}[r_{14}])^2$} \\
        0 & \text{otherwise}
    \end{cases}
\end{align}

\begin{align}\label{eq:fcnot2}
    F^{(2)}_{\text{CNOT}}(\rho) = \begin{cases}
        r_2 & \text{\footnotesize if $(2r_2-1)
  (1-2r_3)>-(2{\rm Im [r_{14}]})^2-(2{\rm Re}[r_{23}])^2 $}\\
        0 & \text{otherwise}.
    \end{cases}
\end{align}

Output fidelities need to be computed according to Eqs. \eqref{eq:fbennett}--\eqref{eq:fcnot2} for each sampled density matrix at each iteration of the corresponding protocol. The average fidelity is defined as

\begin{align}\label{eq:avfid}
    \bar{F}^{(k)}_{\textrm{protocol}}(\rho) =  \frac{1}{N} \sum_{j=1}^N F^{(k)}_{\textrm{protocol}}(\rho_j),
\end{align}
where $k \in \{1,2,4\}$ is the label of the Bell states which are stable fixed points, and $N$ is the sample size.

\begin{figure}
    \hspace{-1cm}
    \centering
    \includegraphics[scale=0.25]{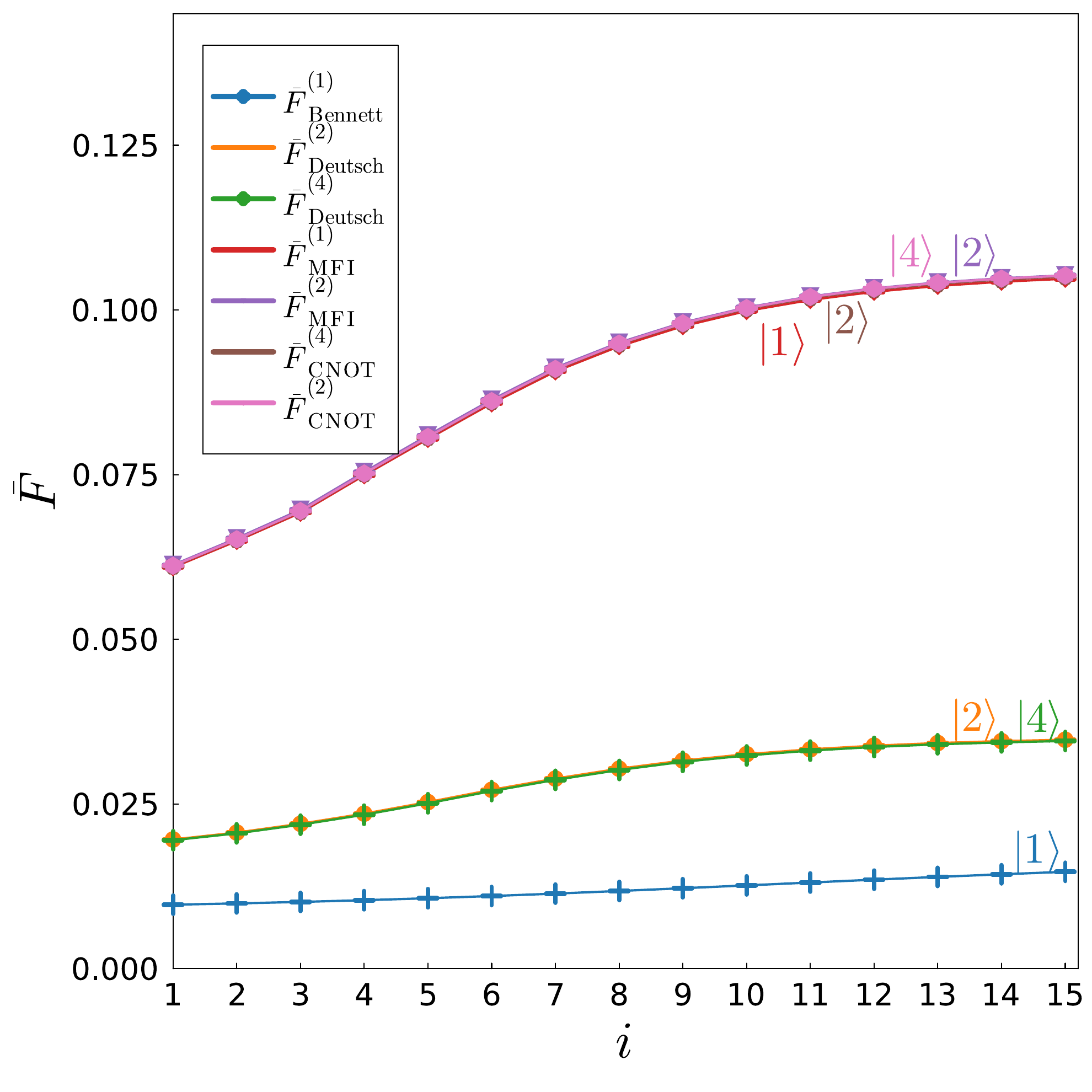}
    \caption{Average output fidelities [calculated using Eqs. \eqref{eq:fbennett}--\eqref{eq:fcnot2}, and \eqref{eq:avfid} for the different fixed points] with respect to the attractors of the different protocols as a function of the number of iterations. For the Bennett protocol, $\ket{1}$ is the only fixed point. For the other protocols, the attractors are $\ket{2}$ and $\ket{4}$ (Deutsch), $\ket{1}$ and $\ket{2}$ (MFI) and $\ket{4}$ and $\ket{2}$ (CNOT). We see that for each of their attractors, the shares of purifiable states of both the MFI and the CNOT protocols are significantly larger than the ones of Bennett and Deutsch. The plot is obtained by running ten simulations of the hit-and-run algorithm, each one with $N=10^6$ samples.}
    \label{fig:fidelities_protocols}
\end{figure}

The average fidelities for the purifiable states with respect to all stable fixed points for the Bennett, Deutsch, MFI, and CNOT protocols are given in Fig.~\ref{fig:fidelities_protocols}.
We see that the number of states that can be brought to a stable fixed point by the MFI and CNOT protocols is significantly larger than those of the Bennett and Deutsch protocols. This is in accordance with the findings presented in the main text.  

Prior knowledge of the input state is essential because most of the protocols currently available in the literature can only work if one knows that the states to be purified have certain underlying properties, which need to be known to avoid failure. In the subsequent discussion, we elucidate this argument by a simple demonstration. Let us now consider a general two-qubit density matrix that we wish to purify towards a Bell state. The Bennett protocol starts with a twirling operation, whose  only goal is to bring the state $\rho$ into the Werner form

\begin{align}
    \rho \overset{\text{twirling}}{\longrightarrow} \rho_{W} = r_1 \ket{1}\bra{1} + \frac{1-r_1}{3}\left(I_4 - \ket{1}\bra{1}\right).
\end{align}

This operation, however, does not guarantee the success of the purification because the Bennett protocol can only work if the condition $r_1 > 0.5$ is fulfilled. If it does, then the Bennett protocol in the asymptotic limit brings this state to the Bell state $\ket{1}$.
In all other cases, the purification protocol fails and leads to a mixed state. The infinite-limit output of the Bennett protocol for a general two-qubit density matrix is
therefore a classical mixture of the  purified states (with concurrence one) and the mixed states for which the protocol failed (with concurrence zero). It
turns out that for most of the density matrices, the protocol fails, i.e. the twirling operation maps them to Werner states with $r_1 < 0.5$, which is why the blue curve in Fig.~\ref{fig:protocolcomparison} has such small values. The asymptotic value of the curve represents exactly the fraction of states that can be purified. The output of the Bennett protocol with asymptote $\bar{\mathcal{C}}^{(\infty)}_{\text{Bennett}}$ for a general random density matrix in the limit of infinite iterations will be approximately
\begin{align}
    \rho_{\text{Bennett}} = \bar{\mathcal{C}}^{(\infty)}_{\text{Bennett}} \ket{1}\bra{1} + \frac{(1-\bar{\mathcal{C}}^{(\infty)}_{\text{Bennett}})}{3}(I_4 - \ket{1}\bra{1})
\end{align}
which is again a Werner state, but with fidelity $\bar{\mathcal{C}}^{(\infty)}_{\text{Bennett}}$. This presents a limitation of using the Bennett protocol on larger classes of states, especially if their generation cannot be controlled in such a way that the condition $r_1 > 0.5$ is met. This implies that for the Bennett protocol information about the input state has to be given otherwise we get a very noisy entangled output state even after infinitely many iterations. 
The Deutsch protocol, unlike the Bennett, has three stable fixed points with two Bell states, which is often not mentioned in the literature \cite{Duer2007}, but after the appearance of the proposal this has been thoroughly investigated in Ref. \cite{Macchiavello}. Now, one obtains a classical mixture of three states with weights defined by $\bar{\mathcal{C}}^{(\infty)}_{\text{Deutsch}}$, where the  states $\ket{2}\bra{2}$ and $\ket{4}\bra{4}$ have the same weights as it is shown in Fig.~\ref{fig:protocolcomparison}. This leads also to a noisy entangled state. In conclusion,  one cannot start with an unknown density matrix and just run the protocols without having some prior information beforehand, because, as we see, the protocols mostly fail if certain conditions are not met. These conditions, as we have shown, may be very restrictive or more relaxed, in which case a broader class of input states can be successfully purified. 

\bibliography{sample}

\onecolumngrid

\end{document}